\documentclass[12pt]{article}
\usepackage{epsfig}

\usepackage{latexsym}
\usepackage{indentfirst}
\usepackage{fancyhdr}
\usepackage{amssymb}
\usepackage{amsmath}
\usepackage{multirow}
\usepackage{hhline}
\usepackage{amsfonts}
\usepackage{pifont}
\usepackage{cite}
\usepackage{color}
\usepackage{slashed}
\usepackage{graphics}
\usepackage{url}
\usepackage{array}
\usepackage{bm}
\usepackage{graphicx}
\usepackage{float}
\usepackage{cleveref}
\usepackage{breqn}

\usepackage{array,multirow}
\usepackage{hhline}
\usepackage{graphicx}
\usepackage{subcaption}

\begin{document}

\begin{titlepage}
 \begin{flushright}
TTP18-031
 \end{flushright}
   \vskip 1cm
   \begin{center}
    {\Large\bf Gauge Coupling Unification without Supersymmetry }
   
   \vskip 0.2  cm
   \vskip 0.5  cm
Jakob Schwichtenberg$^{\,a,}$\footnote{E-mail: \texttt{jakob.schwichtenberg@kit.edu}},
\\[1mm]
   \vskip 0.7cm
 \end{center}

\centerline{$^{a}$ \it  Institut f\"ur Theoretische Teilchenphysik, 
Karlsruhe Institute of Technology,}
\centerline{\it  Engesserstra{\ss}e 7, D-76131 Karlsruhe, Germany} 
\vspace*{1.5cm}

\begin{abstract}
\noindent
We investigate the prospects to achieve unification of the gauge couplings in models without supersymmetry. We restrict our discussion to $SU(5), SO(10)$ and $E_6$ models that mimic the structure of the Standard Model as much as possible ("conservative models"). One possible reason for the non-unification of the standard model gauge couplings are threshold corrections which are necessary when the masses of the superheavy fields are not exactly degenerate. We calculate the threshold corrections in conservative models with a Grand Desert between the electroweak and the unification scale. We argue that only in conservative $E_6$ models the corrections can be sufficiently large to explain the mismatch and, at the same time, yield a long-enough proton lifetime. A second possible reason for the mismatch are particles at an intermediate scale. We therefore also study systematically the impact of additional light scalars, gauge bosons and fermions on the running of the gauge coupling. We argue that for each of these possibilities there is a viable scenario with just one intermediate scale.

\end{abstract}
\end{titlepage}

\section{Introduction}

Although no experimental hints for a Grand Unified Theory (GUT) were observed so far, the general idea remains as an attractive and popular guideline for models beyond the Standard Model (SM). Among the reasons for the popularity of GUTs are that they allow us to understand the quantization of electric charge, the strengths of the SM coupling constants, why neutrinos are so light and quite generically contain all the ingredients needed to explain the baryon asymmetry \cite{Langacker:1980js}. Over the last decades the main focus of most researchers where supersymmetric GUTs, especially after the famous observation that the gauge couplings meet approximately at a common point if supersymmetric particles are present at a low scale, while they do not in the SM \cite{Ellis:1990zq, Amaldi:1991cn, Ellis:1990wk, Giunti:1991ta, Langacker:1991an}.  
Since so far no hints of supersymmetric particles were experimentally observed, there was recently a revival of non-supersymmetric GUTs \cite{Altarelli:2013lla,Bajc:2005zf,Bertolini:2009es,Joshipura:2011nn,Buccella:2012kc,Altarelli:2013aqa,Babu:2015bna}. In such models gauge unification is possible, for example, if an intermediate symmetry between the GUT and the SM symmetry exists \cite{Rajpoot:1980xy, Yasue:1981nd, Gipson:1984aj, Chang:1984qr, Deshpande:1992au, Deshpande:1992em, Bertolini:2009qj}. However, this is only one possibility out of many and our goal here is to discuss systematically the various possibilities to achieve unification of the gauge couplings in scenarios without supersymmetry. 

After a short discussion of gauge unification in a more general context, we focus on the three most popular GUT groups: $SU(5)$, $SO(10)$ and $E_6$. This restriction is necessary since there are, in principle, infinitely many groups that can be used in GUTs. The group $SU(5)$ is the minimal simple group that contains the SM and was the group used in the original proposal by Georgi and Glashow \cite{Georgi:1974sy}. An attractive feature of $SO(10)$ models \cite{Fritzsch:1974nn} is that the fundamental spinor representation not only contains the SM particles but also a right-handed neutrino. This additional neutrino in each generation is, for example, a crucial ingredient to realize the type-I seesaw \cite{Minkowski:1977sc,Mohapatra:1979ia,Ramond:1979ia,Yanagida:1980xy}. Lastly, $E_6$ \cite{Gursey:1975ki} is popular since it is the only exceptional group that can be used without major problems in a conventional GUT. The exceptional status is interesting because, in contrast, $SU(5)$ is part of the infinite $SU(N)$ family, $SO(10)$ of the infinite $SO(N)$ family and "\textit{describing nature by a group taken from an infinite family does raise an obvious question - why this group and not another?}" \cite{Witten:2002ei}. Moreover, the fundamental representation of $E_6$ contains additional exotic fermions which makes it possible to construct $E_6$ models which solve the dark matter or strong CP puzzle \cite{Schwichtenberg:2017xhv,Schwichtenberg:2018aqc}. 

Unfortunately it is not sufficient to specify the GUT group, since with any given group infinitely many different models can be constructed. One reason for this ambiguity is that there is no fundamental principle that fixes the scalar and fermion representations in GUTs. Moreover, for larger groups like $SO(10)$ or $E_6$ there are dozens of different breaking chains from the GUT group down to ${G_{SM} \equiv SU(3)_C\times SU(2)_L \times U(1)_Y}$. Therefore it is necessary that we restrict ourselves to a finite subset of possible scenarios. For this reason, we define a subcategory consisting of all models that mimic the structure of the SM as much as possible. In the following, we call this subcategory "conservative models". Mimicking the structure of the SM exactly would mean for the particle content:

\begin{itemize}
    \item Only scalars that couple to the fermions.  
    \item Only fermions that live in the fundamental or trivial representation of the gauge group.
    \item Only gauge bosons in the adjoint representation. 
\end{itemize}

However, $SU(5)$ and $SO(10)$ scenarios that fulfill these criteria are phenomenologically nonviable and we are therefore forced to add additional representations. Still, we want to stay as closely as possible to the structure of the SM and therefore only add the minimal representations necessary. The fundamental representation of $SU(5)$ is only $5$-dimensional and therefore cannot contain all SM fermions of one generation. Therefore, we have to add an additional fermionic $10$. Moreover, in $SU(5)$ and $SO(10)$ models the scalar representations that couple to the fermions cannot accomplish the breaking down to $G_{SM}$. For this reason we add in both cases a scalar adjoint. These choices can also be understood through the embedding $SU(5) \subset SO(10) \subset E_6$, since $E_6$ models always contain exotic fermions and no additional representations are necessary.

We start in Section~\ref{sec:noncanonicalhypercharge} with a general discussion of the renormalization group equations (RGEs) and the hypercharge normalization. In Section~\ref{sec:thresholdcorrections} we then discuss unification in conservative $SU(5)$, $SO(10)$ and $E_6$ models with a "Grand Desert" between the electroweak and the GUT scale. Afterwards, we discuss the impact of additional light scalars, fermions and gauge bosons on the running of the gauge couplings. Here and in the following "light" always means light when compared to the GUT scale.

\section{The RGEs and hypercharge normalization}
\label{sec:noncanonicalhypercharge}

The RGEs for the gauge couplings up to two-loop order are
\begin{equation} \label{eq:twolooprge}
\frac{d\omega_i(\mu)}{d \ln \mu} = - \frac{a_i}{2 \pi} - \sum_j \frac{b_{ij}}{8  \pi^2 \omega_j}, 
\end{equation}
where the indices $i,j$ denote the various subgroups at the energy scale $\mu$ and
\begin{equation}
\omega_i = \alpha_i^{-1} = \frac{4\pi}{g_i^2}.
\end{equation}
The coefficients $a_i$ and $b_{ij}$ depend on the particle content and can be calculated manually using the formulas in Ref.~\cite{Jones:1981we} or, for example, with the Python tool PyR@TE 2 \cite{Lyonnet:2016xiz}. While these equations together with the boundary conditions \cite{Patrignani:2016xqp}

\begin{align} 
\omega_{1Y}(M_Z)& = 98.3686 \notag \\
\omega_{2L}(M_Z)& = 29.5752 \notag \\
\omega_{3C}(M_Z)& = 8.54482 \notag \\f
M_Z &= 91.1876 \text{ \ GeV}.
\label{eq:rgeboundaries}
\end{align}
are sufficient to calculate the running of the gauge $SU(2)_L$ and $SU(3)_C$ couplings, there is an ambiguity in the running of the hypercharge coupling. This comes about since the SM Lagrangian only depends on the product of the gauge coupling constant $g'$ times the hypercharge operator $Y$. Therefore, we can perform the transformation $\left(g',Y\right)\rightarrow\left(n^{-1}g',nY\right)$ for any $n$ without changing the Lagrangian. The couplings run non-parallel and it is therefore possible to pick a specific $n$ such that $\omega_{3C}$, $\omega_{2L}$ and $\omega_{1Y}$ meet at a common point. Here we define $n$ as the normalization constant relative to the “Standard Model normalization” where the left-handed lepton doublets have hypercharge $-1$ and the left-handed quark doublets hypercharge $1/3$. The boundary value for $\omega_{1Y}(M_Z)$ in Eq.~\ref{eq:rgeboundaries} is given in this particular “Standard Model normalization”. The RGE coefficients in the SM with this normalization of the hypercharge are

\begin{figure}[h]
    \centering
    \includegraphics[width=0.8\textwidth]{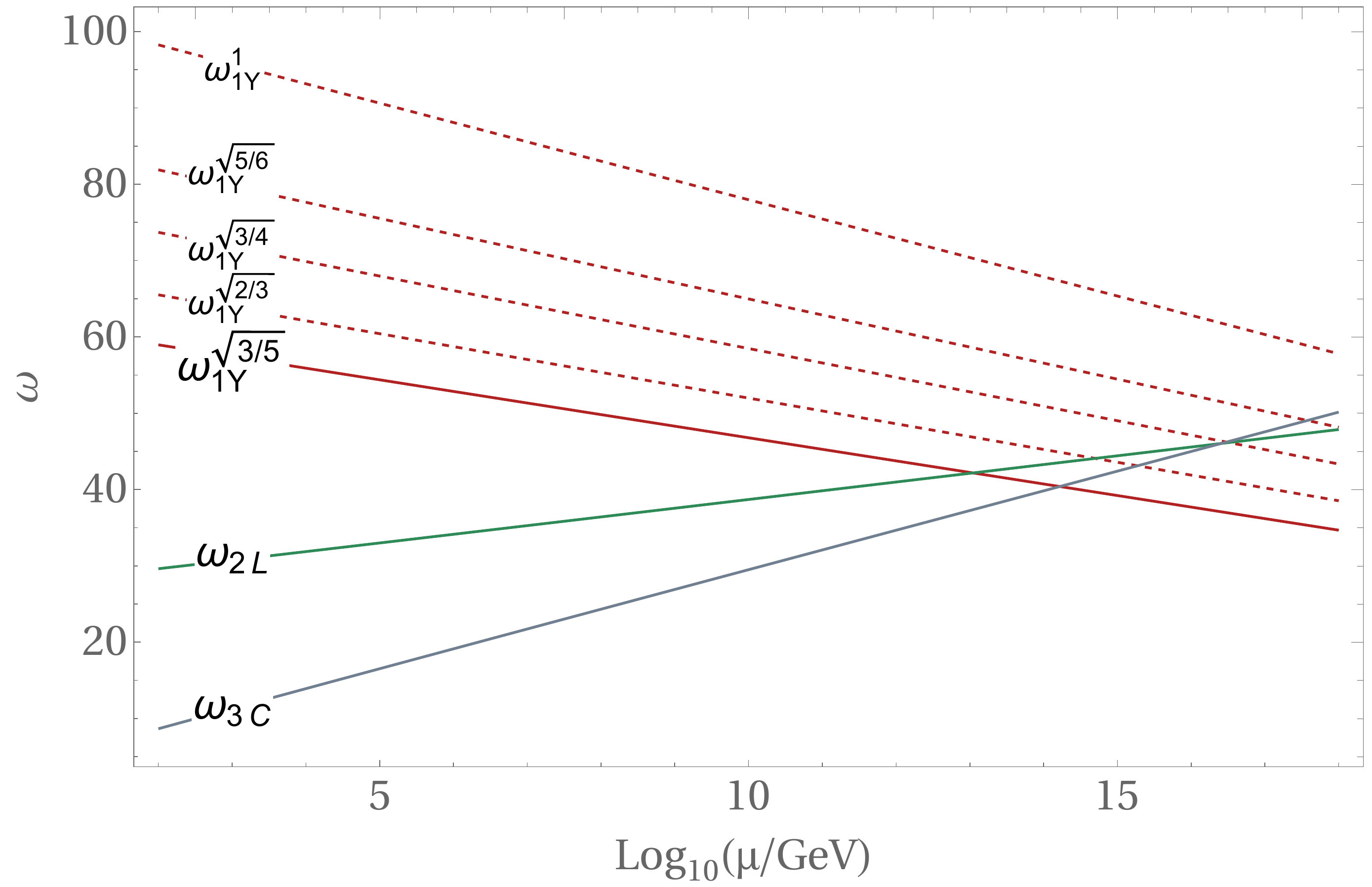}
    \caption{Solutions of the 2-loop RGEs for the Standard Model gauge couplings with different normalizations of the hypercharge, as indicated by the superscripts. The solid line corresponds to the canonical normalization that we get, for example, in $SU(5)$, $SO(10)$ and $E_6$ GUT models. We can see here that with a non-canonical normalization of the hypercharge $n_Y \approx \sqrt{3/4}$ the SM gauge couplings do meet at a point. }
    \label{fig:rgeswithdifferenthyperchargenormalizations}
\end{figure}

\renewcommand\arraystretch{1.2}
\begin{align}
a_{\text{SM}}=\left(
\begin{array}{c}
 \frac{41}{6} , -\frac{19}{6} , -7 
\end{array}
\right) \, , \qquad
b_{\text{SM}}= \left(
\begin{array}{ccc}
 \frac{199}{18} & \frac{9}{2} & \frac{44}{3} \\
 \frac{9}{10} & \frac{35}{6} & 12 \\
 \frac{11}{10} & \frac{9}{2} & -26 \\
\end{array}
\right) \, .
\label{eq:smbetacoefficients}
\end{align}
The coefficients and boundary conditions for different choices of $n$ can be calculated by rescaling the values in Eq.~\ref{eq:rgeboundaries} and Eq.~\ref{eq:smbetacoefficients} appropriately. With this information at hand, we can solve the RGEs for different normalizations of the hypercharge. The results for various normalizations are shown in Figure~\ref{fig:rgeswithdifferenthyperchargenormalizations}.

The choice $n=\sqrt{3/5}$ is known as canonical normalization since it follows automatically when we embed $G_{SM}$ in a simple group $G_{GUT}$ like, for example, $SU(5)$, $SO(10)$ or $E_6$. In such models, $Y$ corresponds to one of the generators of the enlarged gauge group and this fixes the normalization $\textrm{Tr}\left(T_{a}^{2}\right)=\textrm{const.}$ since it must be the same as for all other generators of $G_{GUT}$. For example, in $SU(5)$ models we usually embed the $G_{SM}$ representations $d^{c}=\left(\overline{\mathbf{3}},\mathbf{1},\frac{1}{3}n\right)$
and $L=\left(\mathbf{1},\mathbf{2},-\frac{1}{2}n\right)$ in the fundamental $\overline{5}$. We therefore know that the hypercharge generator reads $Y=n\times\textrm{diag}\left(\frac{1}{3},\frac{1}{3},\frac{1}{3},-\frac{1}{2},-\frac{1}{2}\right)$. We can then fix $n$ by using that equivalently the $SU(2)_{L}$ generators must correspond to $SU(5)$ generators. Therefore, the third generator of $SU(2)_{L}$ is given by  $T_{3L}=\textrm{diag}\left(0,0,0,\frac{1}{2},-\frac{1}{2}\right)$. Using 

\begin{equation}
    \textrm{Tr}\left(T_{3L}^{2}\right) = \frac{1}{2} \stackrel{!}{=} \frac{5}{6} n^2 = \textrm{Tr}\left(Y^{2}\right)
\end{equation}
we can conclude $\left|n\right|=\sqrt{3/5}$. It is clear that for a different choice of $G_{GUT}$ or a different embedding of $G_{SM}$ other values for $n$ are possible \cite{PerezLorenzana:1999tf}. However, the value $n_Y = \sqrt{3/5}$ is quite generic since it follows for all realistic models where the SM is embedded in such a way that we can view it as going through an intermediate $SU(5)$ symmetry: ${G_{GUT} \to SU(5) \to G_{SM}}$ \cite{Fonseca:2015aoa}. While the canonical normalization therefore seems almost inevitable, it is important to keep in mind that a different normalization of the hypercharge could, in principle, lead to successful unification of the gauge couplings, especially when we try to go beyond the standard GUT paradigm \cite{Dienes:1996du}.\footnote{For an interesting alternative proposal which, however, unfortunately does not fix the normalization of the hypercharge see Ref.~\cite{Donoghue:2009fn}.}

In the following sections, we consider unification in explicit $SU(5)$, $SO(10)$ and $E_6$ scenarios and therefore always use $n_Y = \sqrt{3/5}$. Before we can move on we have to define a criterion that tells us when the unification of the gauge couplings is successful in a given model. Through the vacuum expectation value that breaks $G_{GUT}$ the additional GUT gauge bosons get a superheavy mass $m_X$. Therefore "\textit{the gauge couplings at scales much larger than $m_X$ will be approximately equal, because the breaking of the [GUT] gauge symmetry has a negligible effect when all the energies in the process are very large compared to $m_X$. But at energy scales much smaller than $m_X$, the gauge couplings of the $SU(3)$, $SU(2)$, and $U(1)$ subgroups are very different, each running with a $\beta$-function determined by low energy physics."} \cite{Georgi:1994qn} Therefore, naively the unification condition reads
$\omega_{1Y}(M_{GUT}) = \omega_{2L}(M_{GUT}) =  \omega_{3C}(M_{GUT})$. However, it is well known that if we use two-loop RGEs this condition must be refined and threshold corrections can alter it significantly \cite{Dixit:1989ff}. These arise when the masses of the various superheavy particles are not exactly degenerate. The thresholds corrections are small for each individual field, but since there are generically a large number of superheavy particles in GUTs, the individual contributions can add up to non-negligible corrections. In principle it is even possible that threshold corrections are the reason that the SM gauge couplings fail to unify in models with canonical hypercharge normalization. In the following section, we discuss the impact of threshold corrections in various GUT scenarios explicitly. Some GUT gauge bosons mediate proton decay and realistic scenarios are therefore only those where the gauge couplings successfully unify at a scale that is high enough to yield a proton lifetime in agreement with the present experimental bound $\tau_P (p \to e^+ \pi^0) > 1.6 \times 10^{34}$ \cite{Miura:2016krn}. If proton decay is mediated dominantly by the superheavy gauge bosons that are integrated out at the GUT scale this experimental bound implies

\begin{equation} \label{eq:protonboundscale}
    \left( \frac{\omega_G}{45} \right) 10^{2(n_U-15)}> 16.6 \, ,
\end{equation}
where $\omega_G$ denotes the unified gauge coupling. For example, for the typical value $\omega_G = 45$ Eq.~\ref{eq:protonboundscale} yields $n_U>15.6$.

\section{Threshold Corrections}
\label{sec:thresholdcorrections}

In this section we assume that there is a "Grand Desert" between the electroweak and the GUT scale, i.e. no particles at an intermediate mass scale. The threshold corrections, already mentioned above, can be expressed in terms of modified matching conditions \cite{Hall:1980kf}
\begin{equation}
\label{eq:thresholddef}
\omega_i(\mu)=\omega_G(\mu)-\dfrac{\lambda_i(\mu)}{12 \pi} ,
\end{equation}
where
\begin{eqnarray} \label{eq:thresholdformula}
\lambda_i(\mu)&=& \overbrace{\left( C_G-C_i \right)}^{\lambda_i^G} -21 \; \overbrace{Tr\left( t_{iV}^2 \ln \frac{M_V}{\mu}\right)}^{\lambda_i^V} \nonumber\\ 
&&  +\underbrace{Tr \left(t_{iS}^2 P_{GB} \ln \dfrac{M_S}{\mu} \right)}_{\lambda_i^S} + 8 \;\underbrace{Tr \left(t_{iF}^2  \ln \dfrac{M_F}{\mu} \right) }_{\lambda_i^F} .
\end{eqnarray}
Here, $S$, $F$, and $V$ denote the scalars, fermions and vector bosons which are integrated out at the matching scale $\mu$, $t_{iS}$,$t_{iF}$, $t_{iV}$ are the generators of $G_i$ for the various representations, and $C_G$ and $C_i$ are the quadratic Casimir operators for the groups $G$ and $G_i$. $P_{GB}$ is an operator that projects out the Goldstone bosons. The traces of the quadratic generators are known as Dynkin indices and can be found, for example, in Ref. \cite{Slansky:1981yr}. To simplify the notation, we define $\eta_j^a=\ln(\frac{M_j}{\mu})$, where $j$ labels a given multiplet. Moreover, we define the GUT scale as the mass scale of the proton decay mediating gauge bosons. We can then define the following quantities that are independent of the unified gauge coupling $\omega_G(\mu)$ \cite{Ellis:2015jwa}

\begin{equation}
    \Delta \lambda_{ij} (\mu) \equiv   \omega_i(\mu) -\omega_j(\mu) = \lambda_j(\mu) -\lambda_i(\mu) ,
    \label{eq:deltalambdadef}
\end{equation}
for $i,j=1,2,3$, $i \neq j$. These quantities can be evaluated in two ways. Firstly, from the IR perspective by evolving the measured low-energy couplings up to some scale $\mu$. Nonzero $ \Delta \lambda_{ij} (\mu)$ indicate how much the gauge couplings fail to unify. Secondly, we can calculate the $ \Delta \lambda_{ij} (\mu)$ from an UV perspective for any given GUT model. Here, the input needed is the mass spectrum of the superheavy particles. If for a specific GUT model the UV structure yields the values required from the IR input, the gauge couplings successfully unify. In the following, we work with $\Delta \lambda_{12}$ and $\Delta \lambda_{23}$, but any other choice of two $\Delta \lambda_{ij}$ would be equally sufficient.

\begin{figure}[ht]
    \centering
    \includegraphics[width=0.8\textwidth]{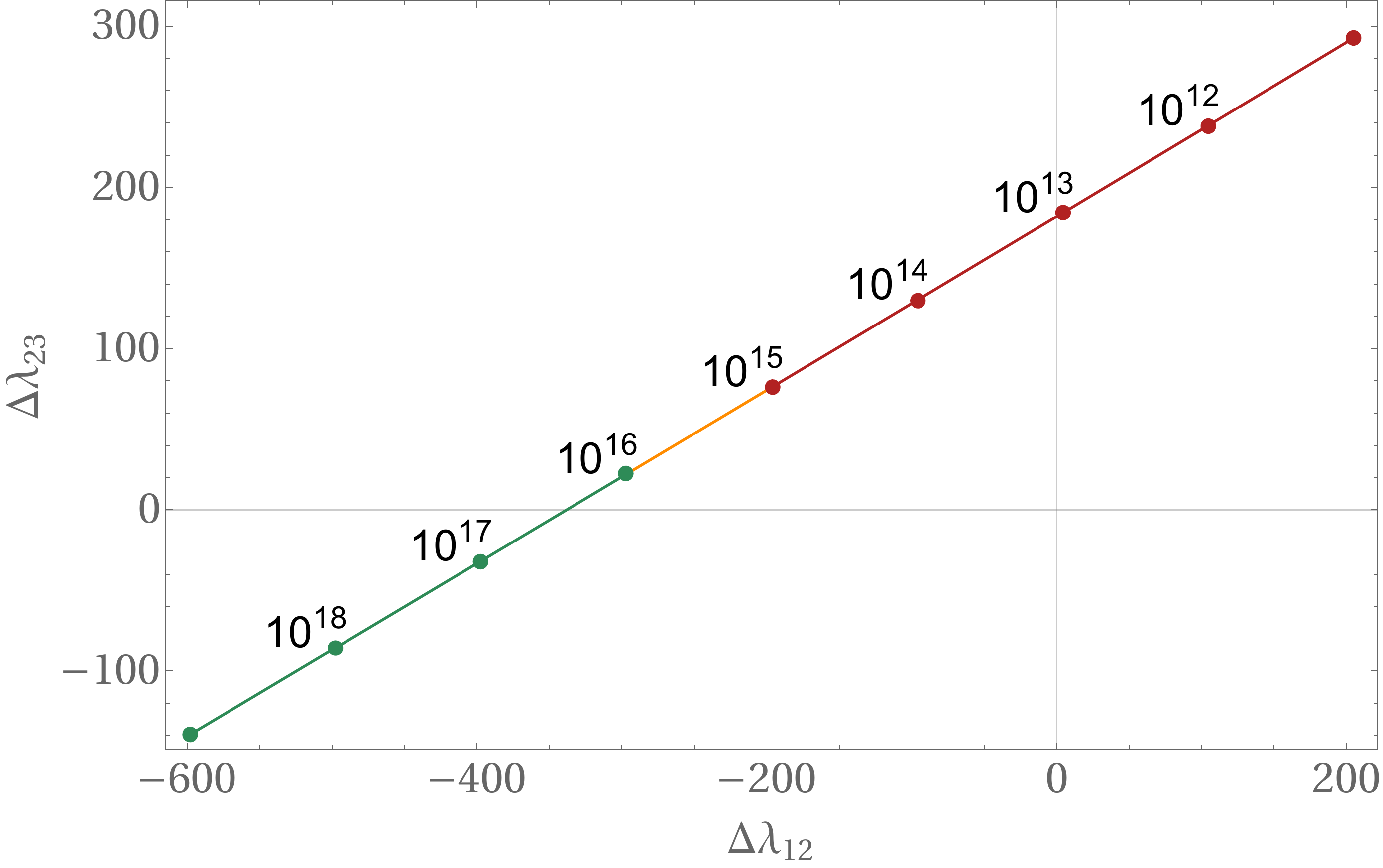}
    \caption{The quantity $\Delta \lambda_{23}(\mu)$ as a function of $\Delta \lambda_{12}(\mu)$ as calculated from the IR input in Eq.~\eqref{eq:rgeboundaries} for GUT models with a Grand Desert between the electroweak and the GUT scale. The quantities $ \Delta \lambda_{ij} (\mu)$ are defined in Eq.~\eqref{eq:deltalambdadef} and indicate how much the gauge couplings fail to unify at a specific scale $\mu$. The numbers above the line denote specific values for $\mu$ in GeV. The red part of the line indicates scales which imply a potentially dangerously short proton lifetime. The orange part implies a proton lifetime close to the present bounds, while the green part indicates a safe proton lifetime. }
    \label{fig:smdeltas}
\end{figure}

Figure~\ref{fig:smdeltas} shows $\Delta \lambda_{23}(\mu)$ over $\Delta \lambda_{12}(\mu)$ for a Grand Desert scenario between the electroweak and the GUT scale, as calculated from the IR input in Eq.~\eqref{eq:rgeboundaries}. In the following sections we investigate if the needed values for $\Delta \lambda_{12}$ and $\Delta \lambda_{23}$ can be realized in specific GUT models. To approximate the threshold corrections in a given GUT model, we choose the masses of the superheavy particles randomly in a given range $R$ around the GUT scale: $M_i = R M_{GUT}$. Previous studies used, for example, $R \in [ \frac{1}{10}, 10]$ in Refs.~\cite{Mohapatra:1992dx,Parida:1987bp} or $R \in [ \frac{1}{10}, 2]$ in Ref.~\cite{Babu:2015bna}. For each randomized spectrum, we can calculate the corresponding $\Delta \lambda_{23}(\mu)$ and $\Delta \lambda_{12}(\mu)$ using Eq.~\eqref{eq:thresholdformula} and Eq.~\eqref{eq:deltalambdadef}.

\subsection{$SU(5)$}

In $SU(5)$ models the SM fermions of one generation live in the $\overline{5}\oplus 10$ representation. It follows from \cite{Slansky:1981yr}
\begin{align}
    5 \times 5 &= 10 \oplus 15 \notag \\
    5 \times \overline{10} &= \overline{5} \oplus \overline{45} \notag\\
    \overline{10} \times \overline{10} &= 5 \oplus 45 \oplus 50
\end{align}
that scalars which yield renormalizable Yukawa terms for the SM fermions live in the ${\overline{5} \oplus 5\oplus 10 \oplus 15 \oplus 45 \oplus \overline{45} \oplus 50}$ representation. In addition, the minimal representation to achieve the breaking of $SU(5)$ to the SM gauge group is the adjoint $24$ representation. For completeness, we investigate the threshold correction if all these representations are present. The decomposition of these representations with respect to the SM gauge group is given in Appendix~\ref{app:su5thresholdsanddecomposition}.

\begin{figure}[ht]
    \centering
    \includegraphics[width=0.8\textwidth]{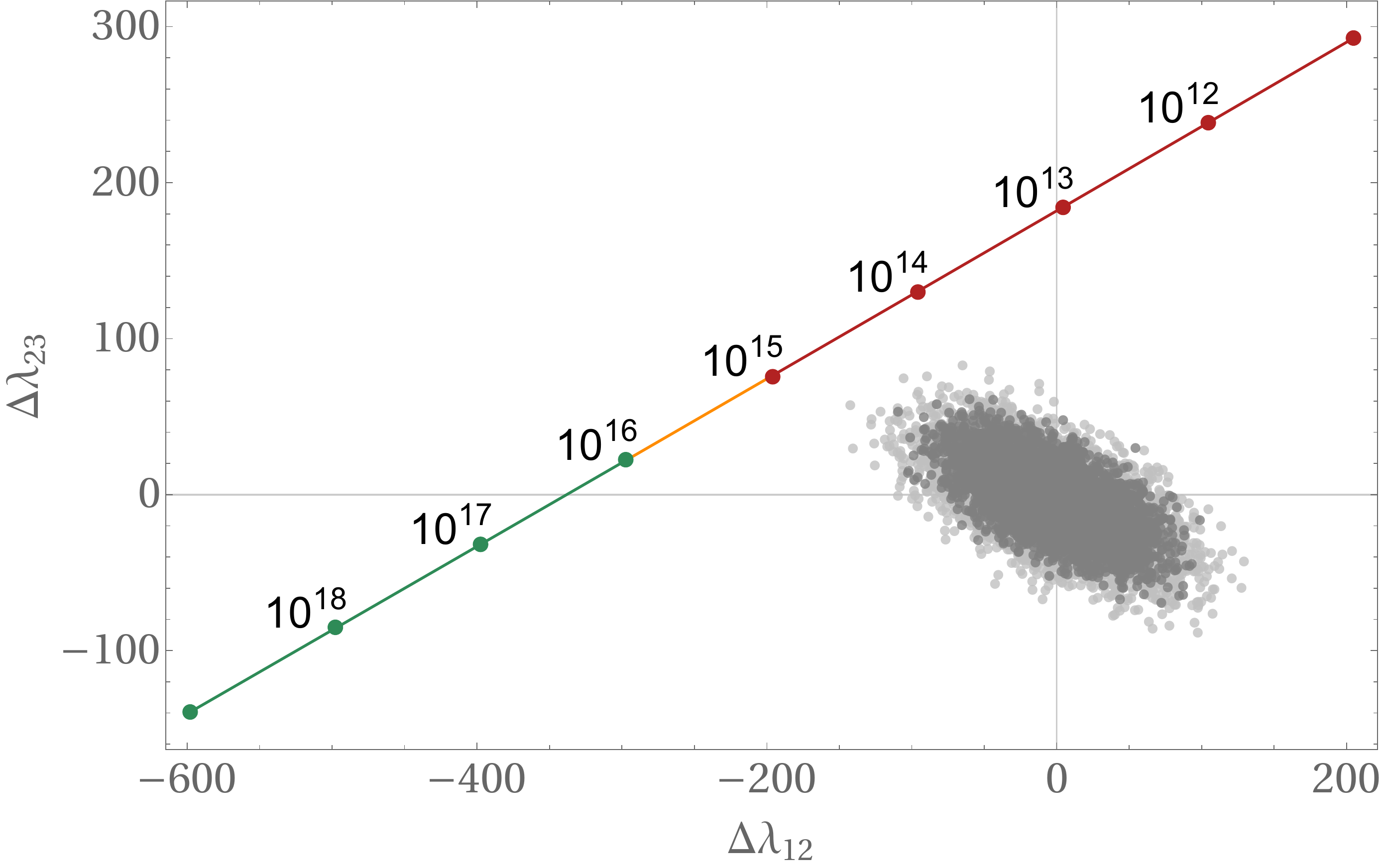}
    \caption{Possible threshold corrections in an $SU(5)$ GUT with scalars in the $5\oplus 10 \oplus 15 \oplus 23\oplus 45 \oplus 50$ representation. The gray points indicate the values for $\Delta \lambda_{23}(\mu)$ over $\Delta \lambda_{12}(\mu)$ for randomized mass spectra of the superheavy particles with $R \in [ \frac{1}{10}, 2]$. The light gray points correspond to $R \in [ \frac{1}{20}, 2]$. Neither with  $R \in [ \frac{1}{10}, 2]$ nor with  $R \in [ \frac{1}{20}, 2]$ configurations exist that could explain the non-unification of the gauge coupling.}
    \label{fig:su5thresh}
\end{figure}

Using Eq.~\eqref{eq:thresholdformula}, we find for this choice of scalar representations
\begin{dgroup*}
\begin{dmath*}
\lambda_{3C} = 2+ \eta_{\varphi_{2}}+\eta_{\varphi_{3}}+\eta_{\varphi_{5}}+2 \eta_{\varphi_{6}}+2 \eta_{\varphi_{8}}+5 \eta_{\varphi_{9}}+3 \eta_{\varphi_{11}}+\eta_{\varphi_{13}}+3 \eta_{\varphi_{14}}+\eta_{\varphi_{15}}+2 \eta_{\varphi_{16}}+5 \eta_{\varphi_{17}}+12 \eta_{\varphi_{18}}+\eta_{\varphi_{20}}+3 \eta_{\varphi_{21}}+\eta_{\varphi_{22}}+2 \eta_{\varphi_{23}}+5 \eta_{\varphi_{24}}+12 \eta_{\varphi_{25}}+\eta_{\varphi_{27}}+2 \eta_{\varphi_{28}}+15 \eta_{\varphi_{29}}+5 \eta_{\varphi_{30}}+12 \eta_{\varphi_{31}} \, , \end{dmath*}
\begin{dmath*}
 \lambda_{2L}  = 3+ \eta_{\varphi_{1}}+3 \eta_{\varphi_{6}}+4 \eta_{\varphi_{7}}+3 \eta_{\varphi_{8}}+2 \eta_{\varphi_{10}}+\eta_{\varphi_{12}}+12 \eta_{\varphi_{14}}+3 \eta_{\varphi_{16}}+8 \eta_{\varphi_{18}}+\eta_{\varphi_{19}}+12 \eta_{\varphi_{21}}+3 \eta_{\varphi_{23}}+8 \eta_{\varphi_{25}}+3 \eta_{\varphi_{28}}+24 \eta_{\varphi_{29}}+8 \eta_{\varphi_{31}} \, ,  
 \end{dmath*}
 \begin{dmath*}
  \lambda_{1Y}  =  5+ \frac{3}{5} \eta_{\varphi_{1}}+\frac{2}{5} \eta_{\varphi_{2}}+\frac{2}{5} \eta_{\varphi_{3}}+\frac{6}{5} \eta_{\varphi_{4}}+\frac{8}{5} \eta_{\varphi_{5}}+\frac{1}{5} \eta_{\varphi_{6}}+\frac{18}{5} \eta_{\varphi_{7}}+\frac{1}{5} \eta_{\varphi_{8}}+\frac{16}{5} \eta_{\varphi_{9}}+\frac{3}{5} \eta_{\varphi_{12}}+\frac{2}{5} \eta_{\varphi_{13}}+\frac{6}{5} \eta_{\varphi_{14}}+\frac{32}{5} \eta_{\varphi_{15}}+\frac{49}{5} \eta_{\varphi_{16}}+\frac{4}{5} \eta_{\varphi_{17}}+\frac{24}{5} \eta_{\varphi_{18}}+\frac{3}{5} \eta_{\varphi_{19}}+\frac{2}{5} \eta_{\varphi_{20}}+\frac{6}{5} \eta_{\varphi_{21}}+\frac{32}{5} \eta_{\varphi_{22}}+\frac{49}{5} \eta_{\varphi_{23}}+\frac{4}{5} \eta_{\varphi_{24}}+\frac{24}{5} \eta_{\varphi_{25}}+\frac{24}{5} \eta_{\varphi_{26}}+\frac{2}{5} \eta_{\varphi_{27}}+\frac{49}{5} \eta_{\varphi_{28}}+\frac{12}{5} \eta_{\varphi_{29}}+\frac{64}{5} \eta_{\varphi_{30}}+\frac{24}{5} \eta_{\varphi_{31}} \, .
\end{dmath*}
\end{dgroup*}
The result of a scan with randomized values of the various masses $M_i = R M_{GUT}$ with $R \in [ \frac{1}{10}, 2]$ or $R \in [ \frac{1}{20}, 2]$ is shown in Figure~\ref{fig:su5thresh}.

We can see that in $SU(5)$ models with a Grand Desert gauge unification cannot be achieved if the masses of the superheavy particles are at most a factor $10$ or $20$ below the GUT scale. 

\subsection{$SO(10)$}

\begin{figure}[ht]
    \centering
    \includegraphics[width=0.8\textwidth]{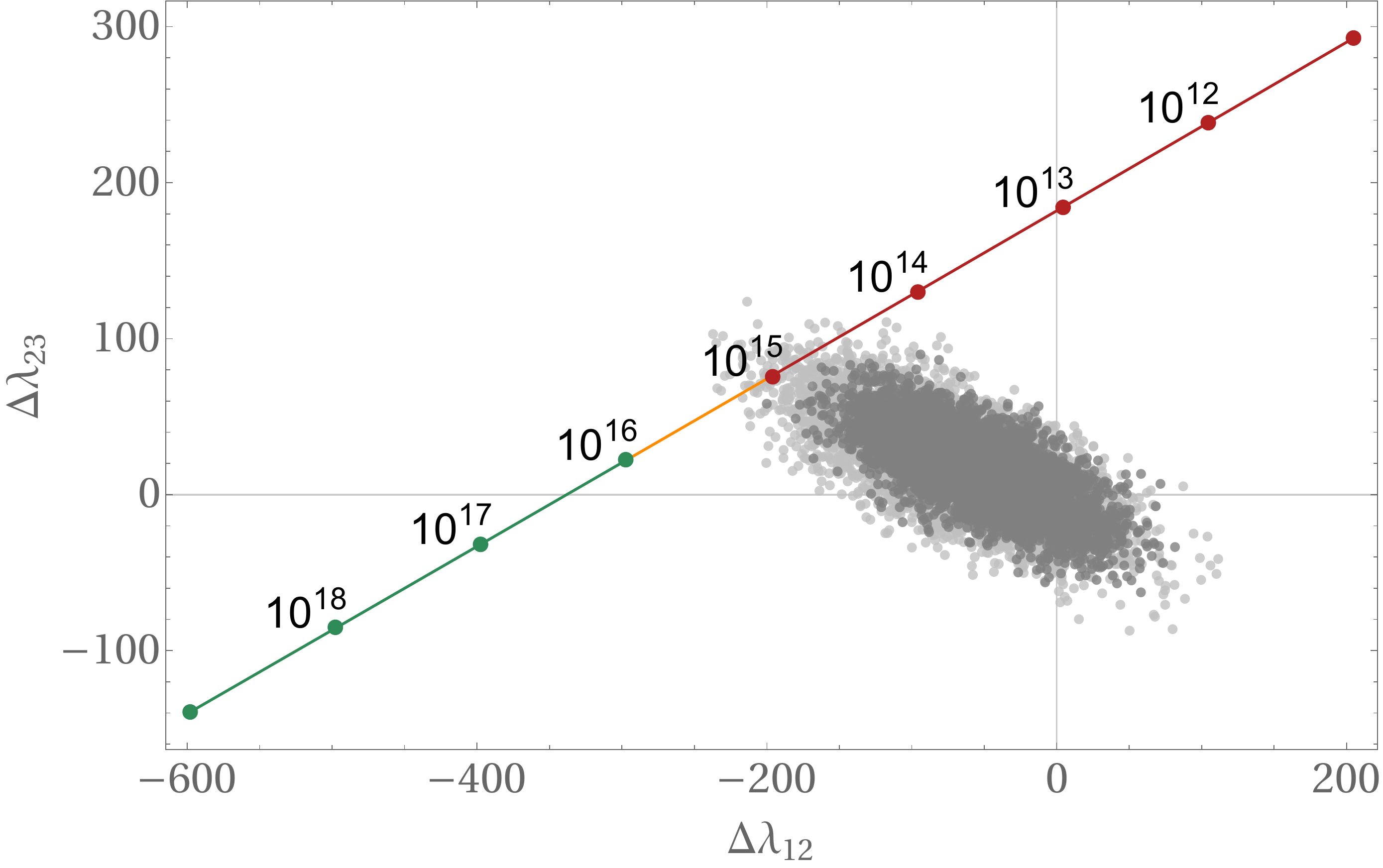}
    \caption{Possible threshold corrections in an $SO(10)$ GUT with scalars in the $10 \oplus 120 \oplus \overline{126} \oplus 45$ representation. The gray points indicate the values for $\Delta \lambda_{23}(\mu)$ over $\Delta \lambda_{12}(\mu)$ for randomized mass spectra of the superheavy particles with $R \in [ \frac{1}{10}, 2]$. The light gray points correspond to $R \in [ \frac{1}{20}, 2]$. While unification is not possible with $R \in [ \frac{1}{10}, 2]$, there are some viable configurations with $R \in [ \frac{1}{20}, 2]$. However, the corresponding maximal $SO(10)$ scale $M_{SO(10)}^{\text{max}} \simeq 10^{15.3}$ GeV, implies a proton lifetime significantly below the present bound (Eq.~\eqref{eq:protonboundscale}).}
    \label{fig:so10thresh}
\end{figure}

In $SO(10)$ models the SM fermions of one generation live in the $16$-dimensional representation. The scalar representations with renormalizable Yukawa couplings to the SM fermions are contained in

\begin{equation}
    \overline{16} \times \overline{16} = 10 \oplus 120 \oplus \overline{126}\, .
\end{equation}
In addition, a $45$ is necessary to break $SO(10)$ down to the SM. Again, for completeness, we consider the threshold effects when all these representations are present. The main difference regarding the threshold corrections, compared to $SU(5)$ models, is that in $SO(10)$ models there are additional gauge bosons which do not mediate proton decay. These do not necessarily have same mass as the proton decay mediating gauge bosons which define the GUT scale. By looking at Eq.~\eqref{eq:thresholdformula} we can see immediately that such additional gauge bosons potentially have a large impact. This is confirmed by a scan with randomized mass of the superheavy fermions $M_i = R M_{GUT}$ with $R \in [ \frac{1}{10}, 2]$ and $R \in [ \frac{1}{20}, 2]$ as shown in Figure~\ref{fig:so10thresh}. The decomposition of the scalar representations and the resulting threshold formulas are given in Appendix~\ref{app:so10thresholdsanddecomposition}. While the threshold corrections can be sufficiently large to explain the mismatch of the gauge couplings, the unification scale is too low to be in agreement with bounds from proton decay experiments (Eq.~\eqref{eq:protonboundscale})\footnote{It is, of course, possible to construct models with larger threshold corrections by including additional scalar representations. See, for example, the model in Ref.~\cite{Lavoura:1993su}.}.

\subsection{$E_6$}

\begin{figure}[ht]
    \centering
   \includegraphics[width=0.8\textwidth]{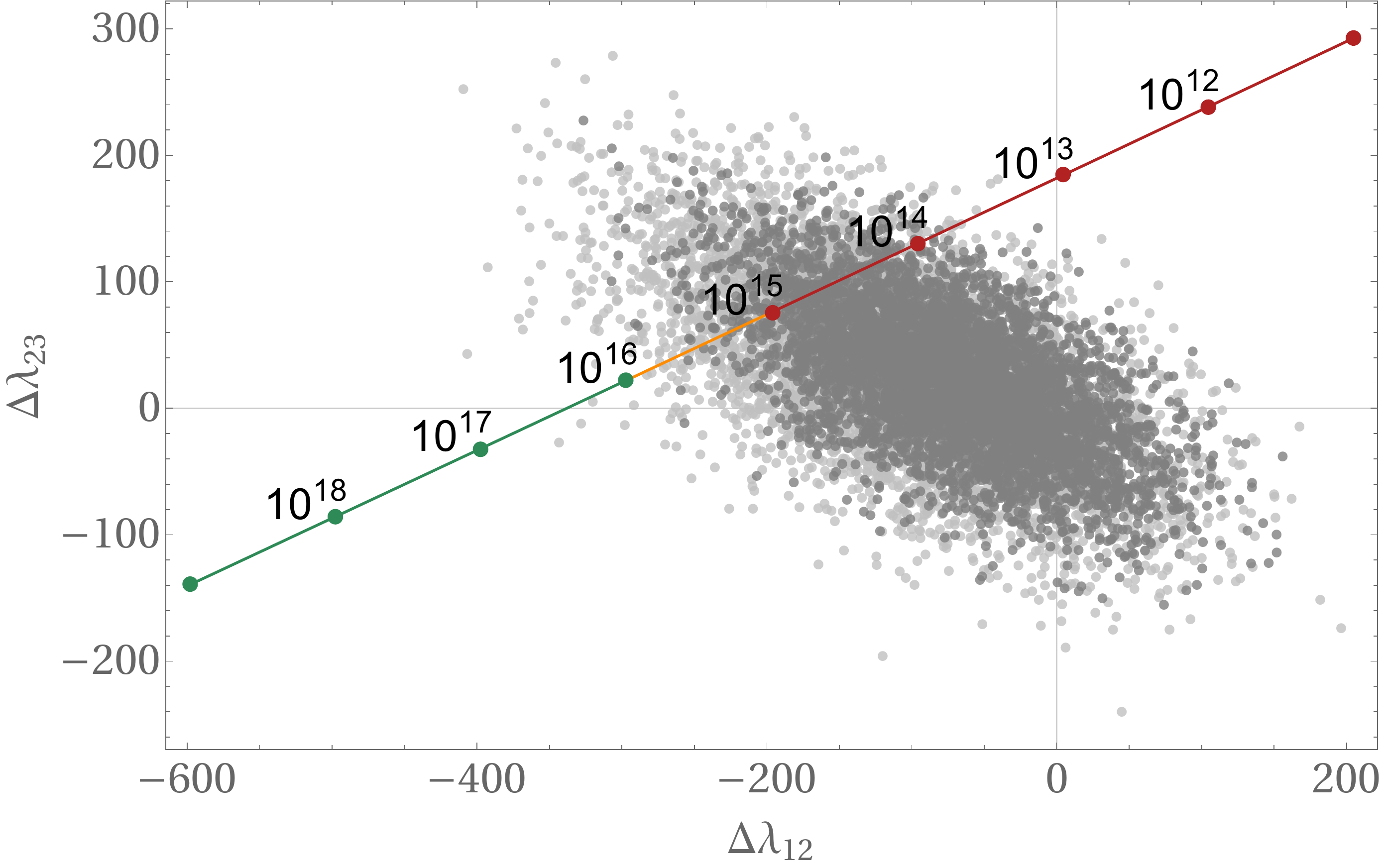}
    \caption{Possible threshold corrections in an $E_6$ GUT with scalars in the $ 27 \oplus 351' \oplus 351$ representation. The $E_6$ scale can be as high as $M_{E_6}^{\text{max}} \simeq 10^{15.8}$ GeV for $R \in [ \frac{1}{10}, 2]$ (gray points) and $M_{E_6}^{\text{max}} \simeq 10^{16.3}$ GeV for $R \in [ \frac{1}{20}, 2]$ (light gray points).}
    \label{fig:e6thresh}
\end{figure}

In $E_6$ models, the SM fermions live in the fundamental $27$-dimensional representation, which decomposes with respect to the maximal subgroup $SO(10) \times U(1)$ as

\begin{equation} \label{eq:27decomposition}
    27 \to 1_4\oplus 10_{-2} \oplus 16_{1} \, .
\end{equation}
The $16_1$ contains, like in $SO(10)$ models, all SM fermions of one generation plus a right-handed neutrino. In addition, we can see that the $27$ contains a sterile neutrino $1_4$ and additionally a vector-like down quark and a vector-like doublet, which are contained in the $10_{-2}$. Since these exotic fermions live in the same representation as the SM fermions, we automatically get $3$ generations of them, too. These additional fermions yield potentially additional significant threshold corrections. The scalars are contained in 

\begin{equation}
    \overline{27} \times \overline{27} = 27 \oplus 351' \oplus 351 \, .
\end{equation}
The decomposition of these scalar representations and the resulting threshold formulas are given in Appendix~\ref{app:e6thresholdsanddecomposition}. In $E_6$, we not only have additional contributions from the three generations of exotic fermions, but also from a larger number of additional gauge bosons and scalars, compared to $SO(10)$ models. Again, we estimate the possible threshold corrections by generating randomized spectra for the superheavy particles. The result is shown in Figure~\ref{fig:e6thresh}. We can see that suitable mass spectra of the large number of superheavy fields in $E_6$ GUTs can indeed explain the mismatch of the gauge couplings.

Next, we investigate whether the non-unification of the gauge couplings could be a hint for particles at intermediate scales. In principle, there can be additional light scalars, fermions and gauge bosons. However, in conservative $SU(5)$ models the only possibility are additional light scalars, while in conservative $SO(10)$ models there can be additional light scalars and gauge bosons, and only in conservative $E_6$ models we can have all three. For this reason, we discuss additional light scalars in the context of $SU(5)$ models, additional light gauge bosons in the context of $SO(10)$ models and additional light fermions in the context of $E_6$ models.

The idea to achieve gauge unification through additional light particles is, of course, not new. For example, to quote E. Ma~\cite{Ma:2005he}: "\textit{If split supersymmetry can be advocated as a means to have gauge-coupling unification as well as dark matter, another plausible scenario is to enlarge judiciously the particle content of the Standard Model to achieve the same goals without supersymmetry.}" Scenarios that realize this idea are discussed extensively in Refs.~\cite{Mambrini:2015vna,Nagata:2015dma}. Our goal here is somewhat different since we are not adding particles solely to achieve gauge unification. Instead, we discuss if it is possible that the gauge couplings meet at a common point with the given particle content in conservative GUTs.

\section{Additional Light Scalars}
\label{sec:addscalars}

Each additional light (non-singlet) particle modifies the RGEs above the scale where it gets integrated out. However, not every modification of the RGEs necessarily brings the gauge couplings closer to unification. A convenient method to check if a given particles improves the running of the gauge couplings was put forward in Ref.~\cite{Giveon:1991zm}. In the following, we use this method and recite here the main points. Firstly, we define the quantities
\begin{equation}
A_{ij}=A_i - A_j \, ,
\end{equation} where
\begin{equation}
\label{r} A_i = a_i+\sum_{I} a_{iI} r_{I}, \qquad r_I=\frac{\ln
M_{GUT}/M_{I}}{\ln M_{GUT}/M_{Z}}.
\end{equation}
Here $a_i$ are the one-loop coefficients as defined in Eq.~\ref{eq:twolooprge}. Necessary (one-loop) conditions for successful gauge unification are then \cite{Giveon:1991zm}

\begin{equation}
\frac{A_{23}}{A_{12}}=\frac{5}{8} \frac{\sin^2
\theta_w-\alpha_{em}/\alpha_s}{3/8-\sin^2 \theta_w},\qquad\qquad
\ln \frac{M_{GUT}}{M_Z}=\frac{16 \pi}{5 \alpha_{EM}}
\frac{3/8-\sin^2 \theta_w}{A_{12}}.
\end{equation}
The left-hand side depends on the particle content, while the experimental input on the right-hand side here is evaluated at $M_Z$. Putting in the experimental values ${\alpha_{EM}^{-1}(M_Z) = 127.950\pm 0.017}$, ${\alpha_{s}(M_Z) = 0.1182\pm 0.0012}$, ${\sin^2 \theta_w(M_Z) = 0.23129 \pm 0.00050}$ \cite{Patrignani:2016xqp} yields
\begin{equation}
\frac{A_{23}}{A_{12}}  \simeq  0.719,\qquad\qquad
\ln \frac{M_{GUT}}{M_Z}  \simeq  
\frac{184.9}{A_{12}}.
\end{equation}
For Grand Desert scenarios, we find $\frac{A_{23}}{A_{12}} \simeq 0.51$. Therefore, a particle brings the gauge couplings closer to gauge unification if it lowers $A_{12}$ and increases $A_{23}$ or if it increases $A_{23}$ more than it increases $A_{12}$. Moreover, from the second relation it follows that particles which lower $A_{12}$ increase the GUT scale. We therefore calculate the contributions to $A_{12}$ and $A_{23}$ for all representations contained in the $5\oplus 10 \oplus 15 \oplus 23\oplus 45 \oplus 50$ representation of $SU(5)$. The result is shown in Table~\ref{table:su5decomp}. We can see that additional light $SU(2)_L$ doublets with the same quantum numbers as the SM Higgs improve the running. However, the contribution is quite small and at least eight of them are needed to bring $\frac{A_{23}}{A_{12}}$ close to the experimental value. Similarly, while helpful, contributions from additional light scalars in the $(1,3,6)$ and $(3,2,1)$ are too small to have a significant impact. The only $SU(3)_C\times SU(2)_L \times U(1)_Y$ representations here with significant impact on the ratio $A_{23}/A_{12}$ are $(1,3,0)$, $(3,3,-2)$ and $(\overline{6},3,-2)$. The RGE coefficients for the SM supplemented with these scalar representations are

\renewcommand\arraystretch{1.2}
\begin{alignat}{2}
&a_{\text{SM}+(1,3,0)}=\left(
\begin{array}{c}
\frac{41}{10},-\frac{5}{2},-7
\end{array}
\right) \, , \qquad
&&b_{\text{SM}+(1,3,0)}= \left(
\begin{array}{ccc}
\frac{199}{50} & \frac{27}{10} & \frac{44}{5} \\
 \frac{9}{10} & \frac{49}{2} & 12 \\
 \frac{11}{10} & \frac{9}{2} & -26 \\
\end{array}
\right) \, , \notag\\
&a_{\text{SM}+(3,3,-2)}=\left(
\begin{array}{c}
\frac{43}{10},-\frac{7}{6},-\frac{13}{2}
\end{array}
\right) \, , \qquad
&&b_{\text{SM}+(3,3,-2)}= \left(
\begin{array}{ccc}
 \frac{207}{50} & \frac{15}{2} & 12 \\
 \frac{5}{2} & \frac{371}{6} & 44 \\
 \frac{3}{2} & \frac{33}{2} & -15 \\
\end{array}
\right) \, , \notag\\
&a_{\text{SM}+(\overline{6},3,-2)}=\left(
\begin{array}{c}
\frac{9}{2},\frac{5}{6},-\frac{9}{2}
\end{array}
\right) \, , \qquad
&&b_{\text{SM}+(\overline{6},3,-2)}= \left(
\begin{array}{ccc}
 \frac{43}{10} & \frac{123}{10} & \frac{124}{5} \\
 \frac{369}{10} & \frac{707}{6} & 172 \\
 \frac{131}{10} & \frac{129}{2} & 89 \\
\end{array}
\right) \, . 
\end{alignat}

    \begin{figure}[ht]
        \centering
        \includegraphics[width=0.8\textwidth]{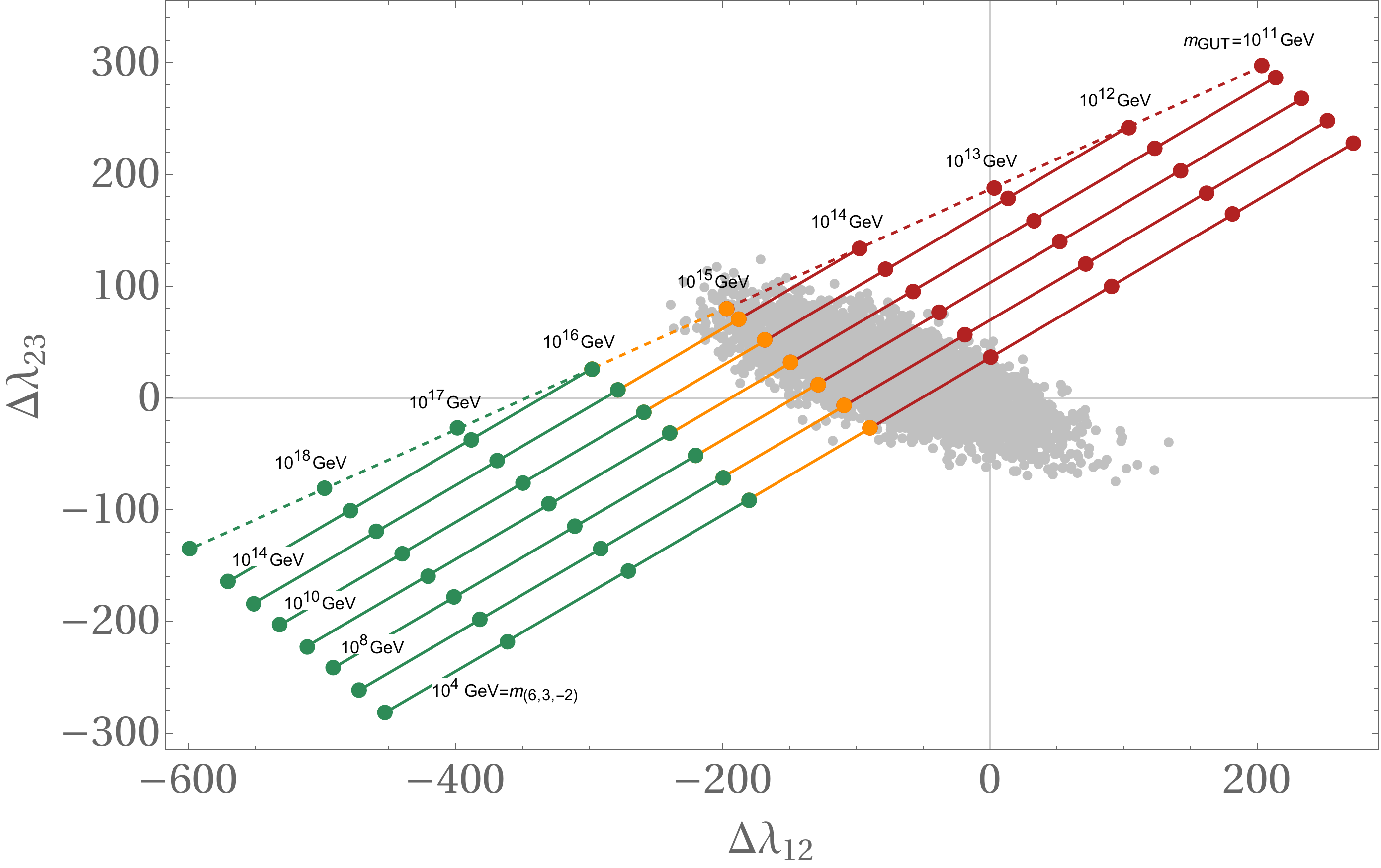}
        \caption{Prospects for gauge unification in scenarios with light $(1,3,0)$ scalars. No sufficiently high unification scale can be realized, even if we take threshold corrections with $R \in [ \frac{1}{20}, 2]$ (light gray points) into account, }
        \label{fig:impactlightscalars3}
    \end{figure}%
    
    \begin{figure}[ht]
        \centering
        \includegraphics[width=0.8\textwidth]{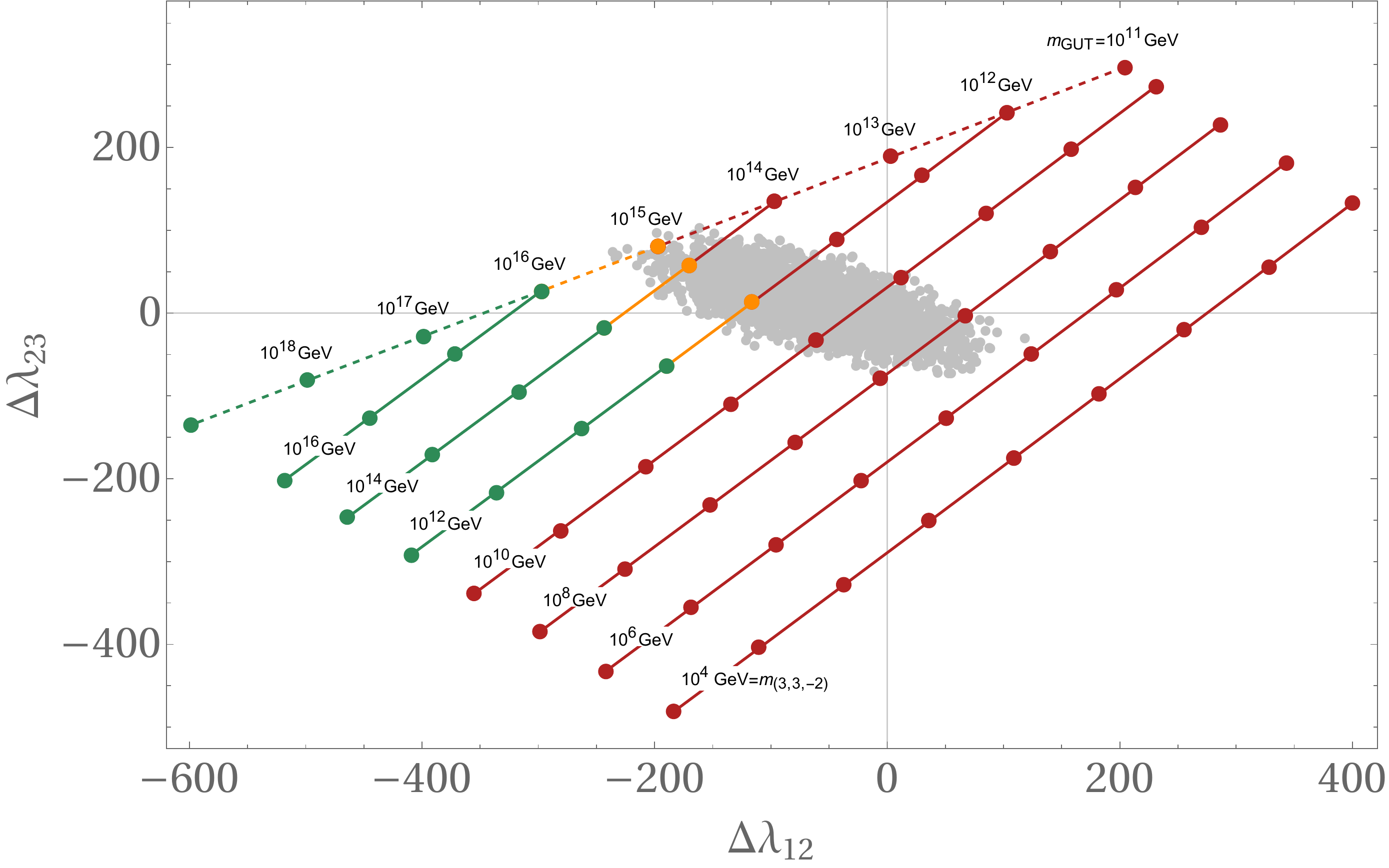}
        \caption{Prospects for gauge unification in scenarios with light $(3,3,-2)$ scalars. Scalars in the $(3,3,-2)$ mediate proton decay and therefore have to be heavier than $10^{10}$~GeV \cite{Dorsner:2006dj}. By taking threshold corrections with $R \in [ \frac{1}{20}, 2]$ (light gray points) into account we find that the $SO(10)$ scale at most $M_{SO(10)}^{\text{max}} \simeq 10^{15.7}$ GeV. This scenario is therefore on the verge of being excluded by proton decay experiments. }
    \end{figure}%

    \begin{figure}[ht]
        \centering
        \includegraphics[width=0.8\textwidth]{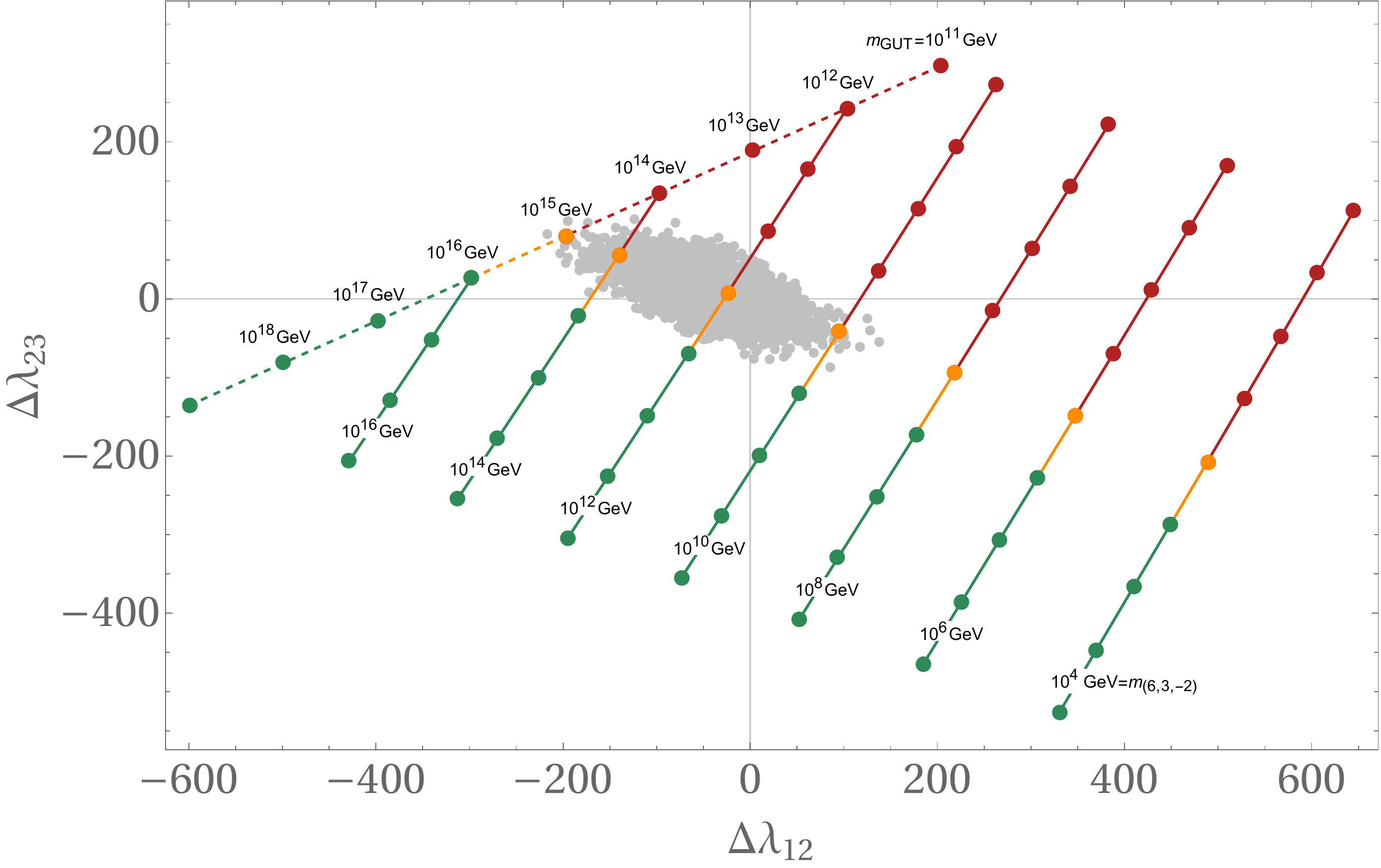}
        \caption{Prospects for gauge unification in scenarios with light $(\overline{6},3,-2)$ scalars. The maximum value for the $SO(10)$ scale $M_{SO(10)}^{\text{max}} \simeq 10^{15.9}$ GeV is possible for ${M_{(\overline{6},3,-2)} \simeq 10^{12} \text{ GeV}}$. This scenario will therefore be probed by the next generation of proton decay experiments \cite{Abe:2011ts}.}
        \label{fig:impactlightscalars1}
    \end{figure}

The impact of these representation on the running of the gauge couplings for various intermediate mass values is shown in Figures~\ref{fig:impactlightscalars3}-\ref{fig:impactlightscalars1}. We can see that no unification at a sufficiently high scale is possible with light $(1,3,0)$ scalars, even if we take threshold corrections into account. The situation is better if there are light $(3,3,-2)$ scalars and the maximum proton lifetime is close to the present bound. For light $(\overline{6},3,-2)$ scalars, the $SO(10)$ scale can be as high as $M_{SO(10)}^{\text{max}} \simeq 10^{15.9}$ GeV if ${M_{(\overline{6},3,-2)} \simeq 10^{12} \text{ GeV}}$. Therefore, this scenario will be probed by the next generation of proton decay experiments \cite{Abe:2011ts}.

Of course, it is also possible to consider scenarios in which more than one scalar representation is light. However, it is well known that each additional light scalar representation requires additional fine-tuning \cite{Mohapatra:1982aq} and since scenarios with just one light representation are still viable, we do not discuss such scenarios any further here.

\section{Additional Light Gauge Bosons}

\label{sec:addgaugebosons}
While in conservative $SU(5)$ models the only possibility to achieve gauge unification is through additional light scalars, in $SO(10)$ and $E_6$ models there can be additionally light gauge bosons, too. This is the case when there is at least one intermediate symmetry between $G_{SM}$ and $G_{GUT}$. Since, the $E_6$ scalar representations that couple to fermions contain no singlet under any viable maximal subgroup other than $SO(10)$, we discuss in the following only breaking chains that start with $G_{GUT}=SO(10)$. Moreover, we restrict ourselves to scenarios with exactly one intermediate symmetry. A thorough discussion of breaking chains with two intermediate symmetries can be found in Ref.~\cite{Bertolini:2009qj}.\footnote{A particularly interesting specific possibility is that $\omega_{1Y}$ and $\omega_{2L}$ unify at around $M_I \simeq 10^{13}$ GeV ($\approx$type-1 seesaw scale) which is where they meet in the SM (c.f. Figure~\ref{fig:rgeswithdifferenthyperchargenormalizations}). A complete unification of the gauge couplings can then be achieved, for example, through additional light scalars \cite{Stech:2008wd}.}

The breaking of $SO(10)$ down to the SM gauge group is achieved by SM singlets in the $10 \oplus 120 \oplus \overline{126} \oplus 45$ scalar representation. There are no SM singlets in the $10$ and $120$ and therefore all superheavy VEVs must come from the $\overline{126}$ or $45$ representation.

The singlet in the $\overline{126}$ breaks $SO(10)$ down to $SU(5)$. Since in such a scenario the gauge couplings already have to unify at the intermediate $SU(5)$ scale there is no improvement compared to the scenarios discussed in the previous section. 

There are two SM singlets in the adjoint $45$ and they can break

\begin{align} 
\label{eq:so10breakingchains}
    SO(10) &\to SU(4)_C \times SU(2)_L  \times U(1)_R \notag \\
    SO(10) &\to SU(3)_C \times SU(2)_L \times SU(2)_R \times U(1)_X \notag \\
    SO(10) &\to SU(3)_C \times SU(2)_L \times U(1)_R \times U(1)_X \notag \\    
    SO(10) &\to SU(5)' \times U(1)_Z  \notag \\
    SO(10) &\to SU(5) \times U(1)_Z  \, ,
\end{align}
Here $SU(5)'$ denotes the flipped $SU(5)$ embedding \cite{DeRujula:1980qc,Barr:1981qv}. The breaking of the intermediate symmetry down to $SU(3)_C \times SU(2)_L \times U(1)_Y$ is achieved for all chains but the last one by the singlet in the $\overline{126}$. For the last chain, the singlet in the $\overline{126}$ only breaks $SU(5) \times U(1)_Z$ down to $SU(5)$. Moreover, the intermediate symmetries $SU(5)' \times U(1)_Z$ and ${SU(3)_C \times SU(2)_L \times U(1)_R \times U(1)_Y}$ do not yield any improvement in terms of unification of the gauge couplings \cite{Deshpande:1992au, Mambrini:2015vna}. There are additional possibilities if we embed $SO(10)$ in $E_6$ since there are additional SM singlets in the $54\subset 351'$ and $144\subset 351$. With a VEV in the $144$ it's possible to break $SO(10)$ directly to $G_{SM}$ \cite{Babu:2005gx} and therefore there is no improvement regarding the running of the gauge couplings. With a VEV in the $54$ we can break $SO(10)$ to the Pati-Salam group ${SU(4)_C\times SU(2)_L \times SU(2)_R \times D}$, where $D$ denotes $D$-parity which exchanges $SU(2)_L \leftrightarrow SU(2)_R$ \cite{Kibble:1982dd, Chang:1983fu}. This breaking chain was analyzed extensively in Refs.~\cite{Babu:2015bna,Chakrabortty:2017mgi}. Hence, in the following we put our focus on the first and second breaking chain in Eq.~\eqref{eq:so10breakingchains}.

Before we can evaluate the RGE running in a scenario with intermediate symmetry, we need to specify the scalar spectrum. For this purpose we use the extended survival hypothesis, which states that "\textit{Higgses acquire the maximum mass compatible with the pattern of symmetry breaking}." \cite{delAguila:1980at}. This a hypothesis of minimal fine tuning since only those scalar fields are light that need to be for the symmetry breaking \cite{Mohapatra:1982aq}. In addition, we need to make sure that the Yukawa sector is rich enough to be able to reproduce the SM fermion observables. For this reason, at least one additional $SU(2)_L$ scalar doublet must be kept at the intermediate scale \cite{Babu:1992ia}.\footnote{In addition to such a minimal choice there is, in general, an extremely large number of alternative possibilities \cite{Deppisch:2017xhv}.}

\subsection{$SO(10) \to SU(4)_C \times SU(2)_L  \times U(1)_R $}

The VEV that breaks $SU(4)_C \times SU(2)_L  \times U(1)_R $ down to the SM gauge group lives in the $(\overline{10},1,-1) \subset \overline{126}$ representation of the intermediate group and therefore has a mass of the order $M_{421}$. The SM Higgs lives in the $(1,2,\frac{1}{2}) \subset 10$ representation. Since at least one additional doublet is needed to generate the flavour structure of the SM, we assume that the $(15,2,\frac{1}{2}) \subset \overline{126}$ has a mass of the order $M_{421}$, too. 

With this particle spectrum, the RGE coefficients above the intermediate scale read

\renewcommand\arraystretch{1.2}
\begin{align}
a_{124}=\left(
\begin{array}{c}
 10,-\frac{2}{3},-7
\end{array}
\right) \, , \qquad
b_{124}= \left(
\begin{array}{ccc}
51 & 24 & \frac{645}{2} \\
 8 & \frac{115}{3} & \frac{285}{2} \\
 \frac{43}{2} & \frac{57}{2} & \frac{265}{2} \\
\end{array}
\right) \, .
\label{eq:smbetacoefficients}
\end{align}
Below $M_{421}$ the RGEs are the Standard Model ones. The matching condition for the hypercharge $U(1)_Y$ without threshold corrections reads\cite{Deshpande:1992au}

\begin{equation}
    \omega_{1Y} = \frac{3}{5} \omega_{1R} + \frac{2}{5} \left( \omega_{4C} - \frac{C_4}{12\pi}\right)
\end{equation}
With this information at hand, we can solve the RGEs and find 

\begin{equation} \label{eq:res421}
    M_{421} \simeq 10^{11.4} \text{ GeV}\, ,\quad M_{SO(10)} \simeq 10^{14.5} \text{ GeV} \, .
\end{equation}
From similar results previous studies concluded that this breaking chain "\textit{is definitely ruled out}" \cite{Deshpande:1992au} since such a low value for $ M_{SO(10)}$ implies a proton lifetime in conflict with experimental bounds. However, as already discussed in Section~\ref{sec:thresholdcorrections}, results such as the one in Eq.~\eqref{eq:res421} can be modified significantly by threshold corrections. 

\begin{figure}[ht]
        \centering
        \includegraphics[width=0.7\textwidth]{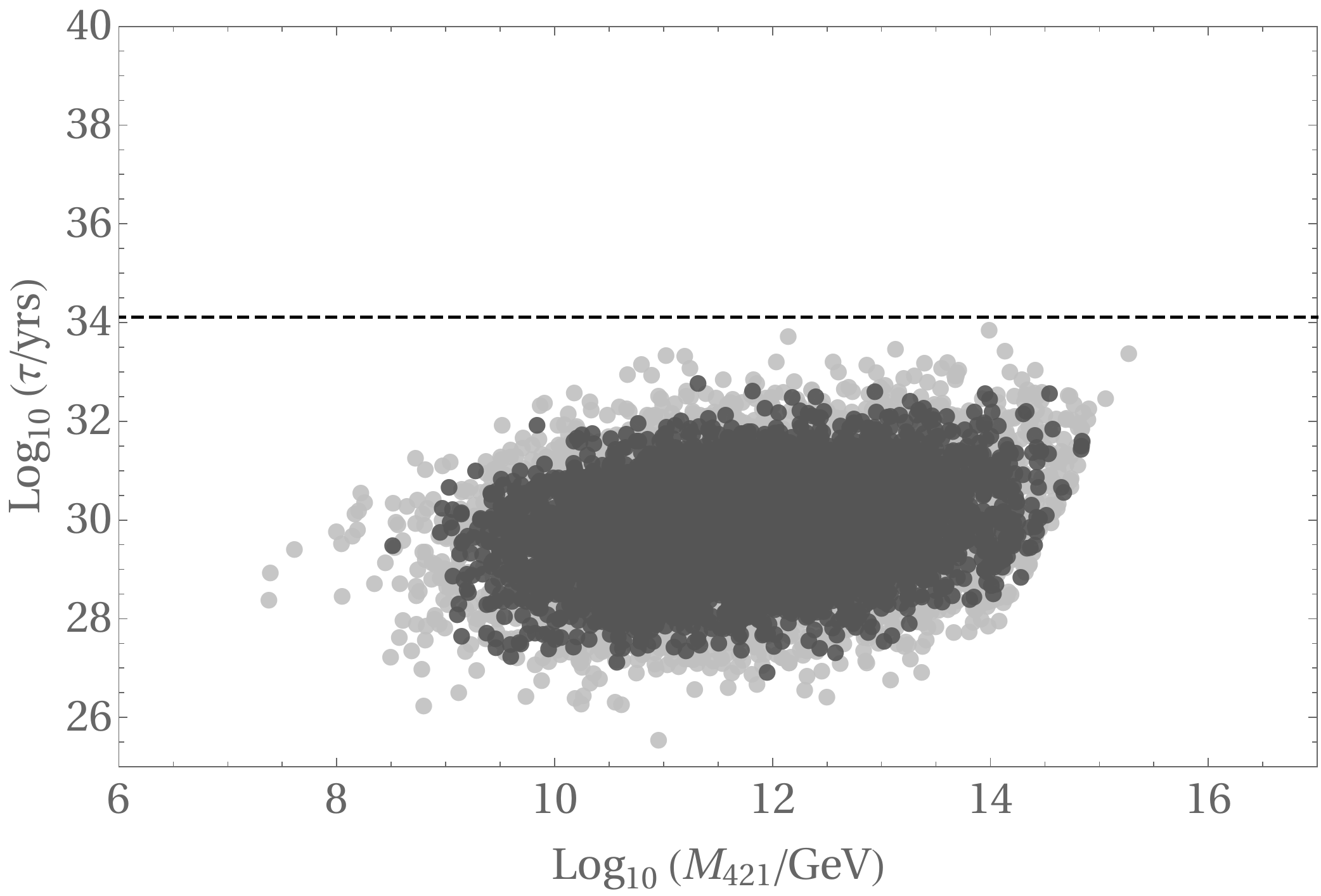}
        \caption{Impact of possible threshold corrections on the proton lifetime $\tau$ in $SO(10)$ models with intermediate $SU(4)_C \times SU(2)_L  \times U(1)_R $ symmetry. The gray dots denote randomized mass spectra with $R \in [ \frac{1}{10}, 2]$ while the light gray dots denote spectra with $R \in [ \frac{1}{20}, 2]$. The dashed line denotes the present bound from Super-Kamiokande \cite{Miura:2016krn}.}
        \label{fig:421thresholds}
\end{figure}%

These depend on the detailed mass spectrum of the superheavy particles and can be estimated by generating the masses of the various multiples randomly $M_i = R M_{j}$, where $j \in \{421 , SO(10)\} $, within a given range, for example, $R \in [ \frac{1}{10}, 2]$ or $R \in [ \frac{1}{20}, 2]$. The decomposition of the relevant scalar representations and the resulting threshold formulas are given in Appendix~\ref{app:so10421thresholdsanddecomposition}. The result of such a scan with randomized mass spectra is shown in Figure~\ref{fig:421thresholds}. We find that within these ranges the proton lifetime can be at most

\begin{align}
    \tau_{max} &=  6.15 \times 10^{32} \text{ yrs.}; \quad R \in [ 1/10, 2] \notag \\
    \tau_{max} &= 7.33 \times 10^{33} \text{ yrs.}; \quad R \in [ 1/20, 2] \, .
\end{align}

We therefore conclude that this breaking chain is ruled out even if we take threshold corrections into account.

\subsection{$SO(10) \to SU(3)_C \times SU(2)_L \times SU(2)_R \times U(1)_X $}

Here, the VEV that breaks the intermediate $SU(3)_C \times SU(2)_L \times SU(2)_R \times U(1)_X$ symmetry lives in the $(1,1,3,-2) \subset \overline{126}$ representation of the intermediate group and the SM Higgs lives in the $(1,2,1,0) \subset 10$ representation. The additional doublet that is needed for the flavour structure of the SM lives in the $(1,2,1,0) \subset \overline{126}$ representation. Therefore, the $(1,2,1,0) \subset 10$ lives at the electroweak scale, the $(1,1,3,1)$ and $(1,2,1,0) \subset \overline{126}$ at the $M_{3221}$ scale, while all other scalars are assumed to be superheavy. 
The RGE coefficients above the intermediate scale read

\begin{figure}[ht]
        \centering
        \includegraphics[width=0.7\textwidth]{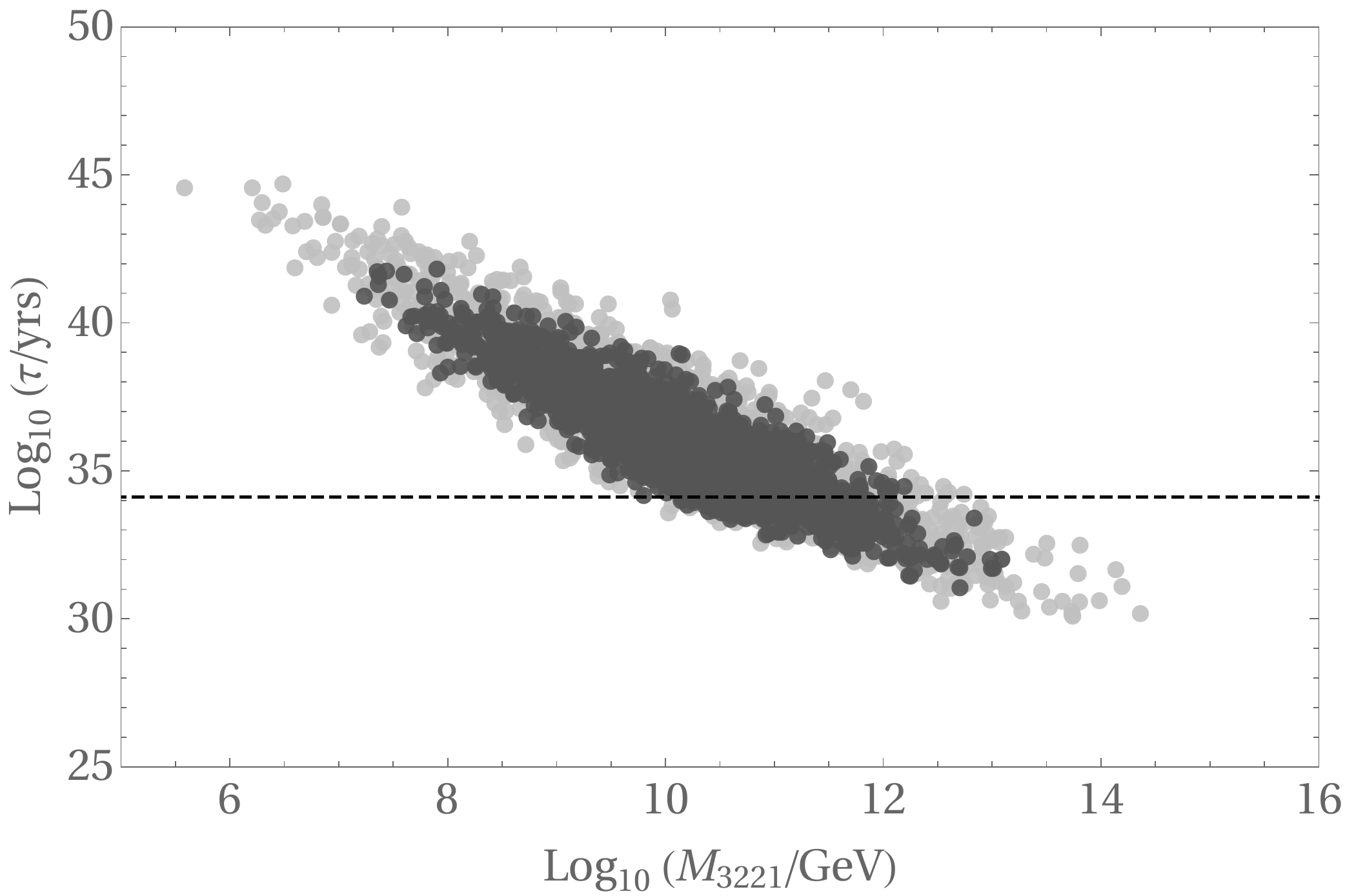}
        \caption{Impact of possible threshold corrections on the proton lifetime $\tau$ in $SO(10)$ models with intermediate $SU(3)_C \times SU(2)_L \times SU(2)_R \times U(1)_X$ symmetry. The gray dots denote randomized mass spectra with $R \in [ \frac{1}{10}, 2]$ while the light gray dots denote spectra with $R \in [ \frac{1}{20}, 2]$. The dashed line denotes the present bound from Super-Kamiokande \cite{Miura:2016krn}.}
        \label{fig:3221thresholds}
\end{figure}

\renewcommand\arraystretch{1.2}
\begin{align}
a_{1223}=\left(
\begin{array}{c}
\frac{11}{2},-\frac{8}{3},-2,-7
\end{array}
\right) \, , \qquad
b_{1223}= \left(
\begin{array}{cccc}
 \frac{61}{2} & \frac{9}{2} & \frac{81}{2} & 4 \\
 \frac{3}{2} & \frac{37}{3} & 6 & 12 \\
 \frac{27}{2} & 6 & 31 & 12 \\
 \frac{1}{2} & \frac{9}{2} & \frac{9}{2} & -26 \\
\end{array}
\right) \, .
\label{eq:smbetacoefficients}
\end{align}
Again, below the intermediate scale the RGEs are the Standard Model RGEs. The matching condition for the hypercharge $U(1)_Y$ without threshold corrections for this breaking chain reads \cite{Deshpande:1992au}

\begin{equation}
    \omega_{1Y} = \frac{3}{5} \left( \omega_{2R}   - \frac{C_2}{12\pi} \right) + \frac{2}{5} \omega_{1X}
\end{equation}
Solving the RGEs yields 

\begin{equation} \label{eq:res421}
    M_{3221} \simeq 10^{10.2} \text{ GeV}\, ,\quad M_{SO(10)} \simeq 10^{15.9} \text{ GeV} \, .
\end{equation}
Therefore, in the absence of threshold corrections this breaking chain is not yet challenged by the experimental bounds on proton decay. Nevertheless, for completeness we investigate the possible impact of threshold corrections. The decomposition of the relevant scalar representations and the resulting threshold formulas are given in Appendix~\ref{app:so103221thresholdsanddecomposition}. The result of a scan with randomized mass of the superheavy particles is shown in Figure~\ref{fig:3221thresholds}. The proton lifetime can be at most

\begin{align}
    \tau_{max} &=  7.16 \times 10^{41} \text{ yrs.}; \quad R \in [ 1/10, 2] \notag \\
    \tau_{max} &= 5.24 \times 10^{44} \text{ yrs.}; \quad R \in [ 1/20, 2] \, .
\end{align}

\section{Additional Light Fermions}

\label{sec:addfermions}

$E_6$ models always contain additional fermions, since the fundamental representation contains in addition to the SM fermions of one generation also exotic fermions. From the decomposition in Eq.~\eqref{eq:27decomposition} it follows that these exotic fermions live in the
\begin{equation} \label{eq:exoticfermions}
    (1,2,3) \oplus (1,2,-3) \oplus (3,1,-2) \oplus (\overline{3},1,2)
\end{equation}
representation of $SU(3)_C \times SU(2)_L \times U(1)_Y$. The additional SM singlets have, of course, no influence on the RGE running. To check which fermions help with gauge unification, we can again use the method discussed in Section~\ref{sec:addscalars}. The contributions of the representations in Eq.~\ref{eq:exoticfermions} to $A_{23}/A_{12}$ are shown in Table~\ref{table:e6fermioncontributions}. We can see here that vector-like lepton doublets $(1,2,3)$ improve the running of the gauge couplings, while vector-like quarks $(3,1,-2)$ make the situation worse. In addition, we can see that at the one-loop level the impact of the vector-like $E_6$ quarks and leptons on the RGE running cancel exactly. 

While the contributions of the individual fermions on the running is quite small, it can be significant since there are three generations of them.\footnote{As already mentioned above, this follows automatically, since they live in the same representation as the SM fermions.} To achieve gauge unification using the exotic $E_6$ fermions, we therefore need a scenario with a large mass splitting between the vector-like leptons and quarks. This is indeed possible since the $45\subset 351$ contains two SM singlets and one of them gives a mass solely to the vector-like quarks, while the other one yields a mass term for the vector-like leptons. Hence, it is possible that the exotic quarks are much heavier than the exotic leptons. This is known as the Dimopoulos-Wilzeck structure \cite{Dimopoulos1983,Srednicki:1982aj}. In the following, we assume that all vector-like quarks are sufficiently heavy to only have a negligible influence on the RGEs and focus solely on the exotic lepton doublet. 

Another crucial observation is that the Yukawa couplings of the exotic fermions and those of the SM fermions have a common origin since the Yukawa sector above the $E_6$ scale reads
\begin{equation}
\label{eq:yukawalagrangiane6}
\mathcal{L}_{\text{Y}}  = \Psi^T i \sigma_2 \Psi (Y_{27} \varphi + Y_{351'} \phi + Y_{351} \xi )  + h.c. \, ,
\end{equation}
It is therefore reasonable to assume that there is a splitting among the three exotic fermion generations which is of comparable size as the splitting among the SM generations, i.e. \\ ${m_{2L}/m_{3L} \simeq 10^{-2}}$, $m_{1L}/m_{3L} \simeq 10^{-4}$. The RGE coefficients for the SM supplemented with one, two and three vector-like lepton doublets are

\renewcommand\arraystretch{1.2}
\begin{align}
&a_{\text{SM+1L}}=\left(
\begin{array}{c}
\frac{9}{2},-\frac{5}{2},-7
\end{array}
\right) \, , \qquad
b_{\text{SM+1L}}= \left(
\begin{array}{ccc}
 \frac{104}{25} & \frac{18}{5} & \frac{44}{5} \\
 \frac{6}{5} & 14 & 12 \\
 \frac{11}{10} & \frac{9}{2} & -26 \\
\end{array}
\right) \, , \notag\\
&a_{\text{SM+2L}}=\left(
\begin{array}{c}
\frac{49}{10},-\frac{11}{6},-7
\end{array}
\right) \, , \qquad
b_{\text{SM+2L}}= \left(
\begin{array}{ccc}
 \frac{217}{50} & \frac{9}{2} & \frac{44}{5} \\
 \frac{3}{2} & \frac{133}{6} & 12 \\
 \frac{11}{10} & \frac{9}{2} & -26 \\
\end{array}
\right) \, , \notag\\
&a_{\text{SM+3L}}=\left(
\begin{array}{c}
\frac{53}{10},-\frac{7}{6},-7
\end{array}
\right) \, , \qquad
b_{\text{SM+3L}}= \left(
\begin{array}{ccc}
 \frac{113}{25} & \frac{27}{5} & \frac{44}{5} \\
 \frac{9}{5} & \frac{91}{3} & 12 \\
 \frac{11}{10} & \frac{9}{2} & -26 \\
\end{array}
\right) \, . 
\end{align}

\begin{figure}[ht]
        \centering
        \includegraphics[width=1.0\textwidth]{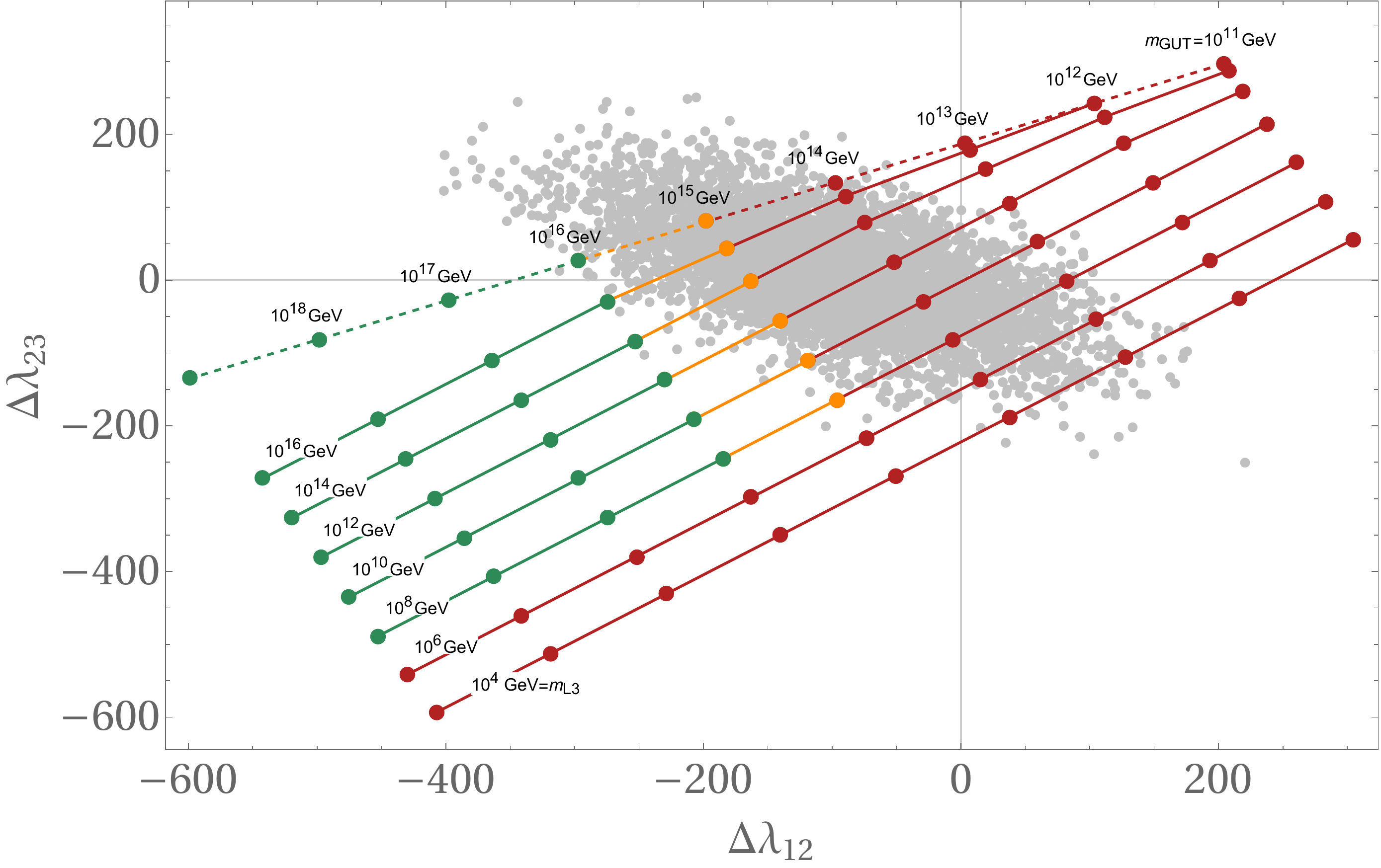}
        \caption{Influence of the 3 generations of exotic $E_6$ lepton doublets $(1,2,3)$ with a mass splitting ${m_{2L}/m_{3L} \simeq 10^{-2}}$, $m_{1L}/m_{3L} \simeq 10^{-4}$ on the unification of the gauge couplings. The numbers in the lower-left corner indicate the mass scale of the heaviest vector-like lepton doublet in each scenario.
        Scenarios with a vector-like lepton doublet lighter than $450$ GeV are already ruled out by collider searches \cite{Falkowski:2013jya}. The dashed line represents the Grand Desert scenario with no particles at intermediate scales. The light gray points indicate possible threshold corrections with $R \in [1/20,2]$. With the heaviest lepton generation around $m_{3L} \simeq 10^{14} \text{ GeV}$, the $E_6$ scale can be as high as 
        $M_{E_6}^{\text{max}} \simeq 10^{16}$ GeV.}
         \label{fig:impactlightfermions}
    \end{figure}

The influence of the vector-like $E_6$ leptons on the running of the gauge couplings is shown in Figure.~\ref{fig:impactlightfermions}. We can see that unification is indeed possible if the mass spectrum of the vector-like leptons is $m_{3L} \simeq 10^{10}$ GeV, $m_{2L} \simeq 10^{8}$ GeV and $m_{1L} \simeq 10^{6}$ GeV. However, the GUT scale in this scenario is dangerously low. 

\section{Summary and Conclusions}

In summary, we have demonstrated that unification of the gauge couplings is possible in conservative GUT scenarios without supersymmetry. 

We have shown that one possible explanation for the observation that the SM gauge couplings do not meet at a common point are large threshold corrections. These are necessary when the superheavy fields do not have exactly degenerate masses. We calculated the magnitude of these corrections in conservative $SU(5)$, $SO(10)$ and $E_6$ models with a Grand Desert between the electroweak and the GUT scale. We found that they can be large enough only in $E_6$ models. The $E_6$ scale can be as high as $M_{E_6}^{\text{max}} \simeq 10^{16.3}$ GeV. 

Afterwards, we investigated scenarios with particles at intermediate mass scales between the electroweak and the GUT scale.

In Section~\ref{sec:addscalars}, we calculated the impact of additional light scalar fields on the running of the gauge couplings. We argued that in conservative $SU(5)$ scenarios the only representations that can significantly help to achieve gauge unification are $(1,3,0)$, $(3,3,-2)$ and $(\overline{6},3,-2)$. While it is possible to achieve unification through suitable mass values for each of these representations (at least if we take threshold corrections into account), only for the $(\overline{6},3,-2)$ this happens at a scale high-enough to be in agreement with bounds from proton decay experiments. In Section~\ref{sec:addgaugebosons}, we investigated scenarios with additional light gauge bosons. In conservative $SO(10)$ GUTs the only scenarios with just one intermediate symmetry and improved running of the gauge couplings go through an ${SU(3)_C \times SU(2)_L \times SU(2)_R \times U(1)_X}$ or ${SU(4) \times SU(2)_L \times U(1)_R}$ stage\footnote{As already mentioned above, in conservative $E_6$ scenarios a third viable possibility goes via $SO(10)$ through an intermediate Pati-Salam symmetry. }. We calculated that the second possibility is already ruled out through proton decay experiments, even if we take threshold corrections into account. For the scenario with ${SU(3)_C \times SU(2)_L \times SU(2)_R \times U(1)_X}$ intermediate symmetry, we found that the proton lifetime can be as long as $\tau_{\text{max}} \simeq 5.24 \times 10^{44} \text{ yrs}$. 

Finally in Section~\ref{sec:addfermions}, we discussed the impact of additional light fermions in the context of conservative $E_6$ models. We argued that light vector-like $E_6$ leptons improve the running, while the vector-like $E_6$ quarks make the situation worse. Including threshold corrections plus the heaviest lepton generation around $m_{3L} \simeq 10^{14} \text{ GeV}$ (and mass splittings ${m_{2L}/m_{3L} \simeq 10^{-2}}$, $m_{1L}/m_{3L} \simeq 10^{-4}$), we found that the $E_6$ scale can be as high as $M_{E_6}^{\text{max}} \simeq 10^{16}$ GeV.

\section*{Acknowledgments}

The author wishes to thank Ulrich Nierste and Paul Tremper for helpful discussions and acknowledges the
support by the DFG-funded Doctoral School KSETA.

\clearpage

\appendix

\section{Decomposition of the Scalar Representations and Threshold Formulas in Grand Desert Scenarios}

\subsection{$SU(5)$}

\label{app:su5thresholdsanddecomposition}

\renewcommand\arraystretch{1.5}
\begin{table}[ph]
\centering
\begin{center}
\tiny
\begin{tabular}{|c|c|c|c|c|}
\hline
$\bm{SU(5})$ & $\bm{SU(3)_C \times SU(2)_L \times U(1)_Y}$ & $\bm{A_{23}/r_I}$ & $\bm{A_{12}/r_I}$ & \textbf{Label} \\
\hline
\multirow{2}{*}{$5$} & $(1,2,3)$ & $\frac{1}{6}$ & -$\frac{1}{15}$ & $\varphi_{1}$ \\
 & $(3,1,-2)$ & -$\frac{1}{6}$ & $\frac{1}{15}$ & $\varphi_{2}$ \\
\hline
\multirow{2}{*}{$\overline{5}$} & $(1,2,-3)$ & $\frac{1}{6}$ & -$\frac{1}{15}$ & $H$ \\
 & $(\overline{3},1,2)$ & -$\frac{1}{6}$ & $\frac{1}{15}$ & $\varphi_{3}$ \\
\hline
\multirow{3}{*}{$10$} & $(1,1,6)$ & $0$ & $\frac{1}{5}$ & $\varphi_{4}$ \\
 & $(\overline{3},1,-4)$ & -$\frac{1}{6}$ & $\frac{4}{15}$ & $\varphi_{5}$ \\
 & $(3,2,1)$ & $\frac{1}{6}$ & -$\frac{7}{15} $ & $\varphi_{6}$ \\
\hline
\multirow{2}{*}{$15$} & $(1,3,6)$ & $\frac{2}{3}$ & -$\frac{1}{15}$ & $\varphi_{7}$ \\
 & $(3,2,1)$ & $\frac{1}{6}$ & -$\frac{7}{15}$ & $\varphi_{8}$ \\
 & $(6,1,-4)$ & -$\frac{5}{6}$ & $\frac{8}{15}$ & $\varphi_{9}$ \\
\hline
\multirow{5}{*}{$24$} & $(1,1,0)$ & $0$ & $0$ & $s_{1}$ \\
 & $(1,3,0)$ & $\frac{1}{3}$ & - $\frac{1}{3}$ & $\varphi_{10}$ \\
 & $(3,2,-5)$ & $\frac{1}{12}$ & $\frac{1}{6} $ & $\xi_{1}$ \\
 & $(\overline{3},2,5)$ & $\frac{1}{12}$ & $\frac{1}{6} $ & $\xi_{2}$ \\
 & $(8,1,0)$ & $-\frac{1}{2}$ & $0$ & $\varphi_{11}$ \\
\hline
\multirow{7}{*}{$45$} & $(1,2,3)$ & $\frac{1}{6}$ & -$\frac{1}{15}$ & $\varphi_{12}$ \\
 & $(3,1,-2)$ & -$\frac{1}{6}$ & $\frac{1}{15}$ & $\varphi_{13}$ \\
 & $(3,3,-2)$ & $\frac{3}{2}$ & -$\frac{9}{5}$ & $\varphi_{14}$ \\
 & $(\overline{3},1,8)$ & -$\frac{1}{6}$ & $\frac{16}{15}$ & $\varphi_{15}$ \\
 & $(\overline{3},2,-7)$ & $\frac{1}{6}$ & $\frac{17}{15}$ & $\varphi_{16}$ \\
 & $(\overline{6},1,-2)$ & $-\frac{5}{6}$ & $\frac{2}{15}$ & $\varphi_{17}$ \\
 & $(8,2,3)$ & -$\frac{2}{3}$ & -$\frac{8}{15}$ & $\varphi_{18}$ \\
\hline 
\multirow{7}{*}{$\overline{45}$}  & $(1,2,-3)$ & $\frac{1}{6}$ & -$\frac{1}{15}$ & $\varphi_{19}$ \\
 & $(\overline{3},1,2)$ & -$\frac{1}{6}$ & $\frac{1}{15}$ & $\varphi_{20}$ \\
 & $(\overline{3},3,2)$ & $\frac{3}{2}$ & -$\frac{9}{5}$ & $\varphi_{21}$ \\
 & $(3,1,-8)$ & -$\frac{1}{6}$ & $\frac{16}{15}$ & $\varphi_{22}$ \\
 & $(3,2,7)$ & $\frac{1}{6}$ & $\frac{17}{15}$ & $\varphi_{23}$ \\
 & $(6,1,2)$ & $\frac{5}{6}$ & $\frac{2}{15}$ & $\varphi_{24}$ \\
 & $(8,2,-3)$ & -$\frac{2}{3}$ & -$\frac{8}{15}$ & $\varphi_{25}$ \\
\hline
\multirow{6}{*}{$\overline{50}$} & $(1,1,-12)$ & $0$ & $\frac{4}{5} $ & $\varphi_{26}$ \\
 & $(3,1,-2)$ & -$\frac{1}{6}$ & $\frac{1}{15}$ & $\varphi_{27}$ \\
 & $(\overline{3},2,-7)$ & $\frac{1}{6}$ & $\frac{17}{15} $ & $\varphi_{28}$ \\
 & $(\overline{6},3,-2)$ & $\frac{3}{2}$ & -$\frac{18}{5}$ & $\varphi_{29}$ \\
 & $(6,1,8)$ & -$\frac{5}{6}$ & $\frac{32}{15}$ & $\varphi_{30}$ \\
 & $(8,2,3)$ & -$\frac{2}{3}$ & -$\frac{8}{15}$ & $\varphi_{31}$ \\
\hline
\end{tabular}
\end{center}

\caption{Decomposition with respect to $SU(3)_C \times SU(2)_L \times U(1)_Y$ of the scalar representations in conservative $SU(5)$ GUTs. Goldstone bosons are labelled by $\xi_i$,  SM singlets by $s_i$ and all other fields by $\varphi_i$. The hypercharges are given in the normalization of Ref.~\cite{Slansky:1981yr}. The numbers in the $A_{23}$ and $A_{12}$ columns indicate whether the fields can help to achieve gauge unification or not. For further explanations, see Section \ref{sec:addscalars}. }
\label{table:su5decomp}
\end{table}

\clearpage

\subsection{$SO(10)$}
  \label{app:so10thresholdsanddecomposition}
Using Eq.~\eqref{eq:thresholdformula}, we find for the threshold corrections in conservative $SO(10)$ GUTs

\begin{dgroup*}
\begin{dmath*}
\lambda_{3C} =  5-21 \eta_{PSV} + \frac{1}{2} \eta_{\Phi_{2}}+\frac{1}{2} \eta_{\Phi_{3}}+\frac{1}{2} \eta_{\Phi_{5}}+\frac{1}{2} \eta_{\Phi_{7}}+\frac{1}{2} \eta_{\Phi_{9}}+\eta_{\Phi_{10}}+\frac{1}{2} \eta_{\Phi_{12}}+\eta_{\Phi_{13}}+\frac{1}{2} \eta_{\Phi_{15}}+\frac{3}{2} \eta_{\Phi_{16}}+\frac{1}{2} \eta_{\Phi_{17}}+\eta_{\Phi_{18}}+\frac{5}{2} \eta_{\Phi_{19}}+6 \eta_{\Phi_{20}}+\frac{1}{2} \eta_{\Phi_{22}}+\frac{3}{2} \eta_{\Phi_{23}}+\frac{1}{2} \eta_{\Phi_{24}}+\eta_{\Phi_{25}}+\frac{5}{2} \eta_{\Phi_{26}}+6 \eta_{\Phi_{27}}+\eta_{\Phi_{29}}+2 \eta_{\Phi_{31}}+5 \eta_{\Phi_{32}}+\eta_{\Phi_{34}}+3 \eta_{\Phi_{35}}+\eta_{\Phi_{36}}+2 \eta_{\Phi_{37}}+5 \eta_{\Phi_{38}}+12 \eta_{\Phi_{39}}+\eta_{\Phi_{41}}+2 \eta_{\Phi_{42}}+15 \eta_{\Phi_{43}}+5 \eta_{\Phi_{44}}+12 \eta_{\Phi_{45}}+\frac{1}{2} \eta_{\Phi_{47}}+\eta_{\Phi_{48}}+\frac{1}{2} \eta_{\Phi_{50}}+\eta_{\Phi_{51}}+3 \eta_{\Phi_{53}} \, , \end{dmath*}
\begin{dmath*}
 \lambda_{2L}  = 6+\frac{1}{2} \eta_{\Phi_{1}}+\frac{1}{2} \eta_{\Phi_{4}}+\frac{1}{2} \eta_{\Phi_{6}}+\frac{3}{2} \eta_{\Phi_{10}}+\frac{3}{2} \eta_{\Phi_{13}}+\frac{1}{2} \eta_{\Phi_{14}}+6 \eta_{\Phi_{16}}+\frac{3}{2} \eta_{\Phi_{18}}+4 \eta_{\Phi_{20}}+\frac{1}{2} \eta_{\Phi_{21}}+6 \eta_{\Phi_{23}}+\frac{3}{2} \eta_{\Phi_{25}}+4 \eta_{\Phi_{27}}+\eta_{\Phi_{28}}+4 \eta_{\Phi_{30}}+3 \eta_{\Phi_{31}}+\eta_{\Phi_{33}}+12 \eta_{\Phi_{35}}+3 \eta_{\Phi_{37}}+8 \eta_{\Phi_{39}}+3 \eta_{\Phi_{42}}+24 \eta_{\Phi_{43}}+8 \eta_{\Phi_{45}}+\frac{3}{2} \eta_{\Phi_{48}}+\frac{3}{2} \eta_{\Phi_{51}}+2 \eta_{\Phi_{52}}\, ,  
 \end{dmath*}
 \begin{dmath*}
  \lambda_{1Y}  = 8-21 \left(\frac{8}{5} \eta_{PSV}+\frac{6}{5} \eta_{W_R}\right)+\frac{3}{10} \eta_{\Phi_{1}}+\frac{1}{5} \eta_{\Phi_{2}}+\frac{1}{5} \eta_{\Phi_{3}}+\frac{3}{10} \eta_{\Phi_{4}}+\frac{1}{5} \eta_{\Phi_{5}}+\frac{3}{10} \eta_{\Phi_{6}}+\frac{1}{5} \eta_{\Phi_{7}}+\frac{3}{5} \eta_{\Phi_{8}}+\frac{4}{5} \eta_{\Phi_{9}}+\frac{1}{10} \eta_{\Phi_{10}}+\frac{3}{5} \eta_{\Phi_{11}}+\frac{4}{5} \eta_{\Phi_{12}}+\frac{1}{10} \eta_{\Phi_{13}}+\frac{3}{10} \eta_{\Phi_{14}}+\frac{1}{5} \eta_{\Phi_{15}}+\frac{3}{5} \eta_{\Phi_{16}}+\frac{16}{5} \eta_{\Phi_{17}}+\frac{49}{10} \eta_{\Phi_{18}}+\frac{2}{5} \eta_{\Phi_{19}}+\frac{12}{5} \eta_{\Phi_{20}}+\frac{3}{10} \eta_{\Phi_{21}}+\frac{1}{5} \eta_{\Phi_{22}}+\frac{3}{5} \eta_{\Phi_{23}}+\frac{16}{5} \eta_{\Phi_{24}}+\frac{49}{10} \eta_{\Phi_{25}}+\frac{2}{5} \eta_{\Phi_{26}}+\frac{12}{5} \eta_{\Phi_{27}}+\frac{3}{5} \eta_{\Phi_{28}}+\frac{2}{5} \eta_{\Phi_{29}}+\frac{18}{5} \eta_{\Phi_{30}}+\frac{1}{5} \eta_{\Phi_{31}}+\frac{16}{5} \eta_{\Phi_{32}}+\frac{3}{5} \eta_{\Phi_{33}}+\frac{2}{5} \eta_{\Phi_{34}}+\frac{6}{5} \eta_{\Phi_{35}}+\frac{32}{5} \eta_{\Phi_{36}}+\frac{49}{5} \eta_{\Phi_{37}}+\frac{4}{5} \eta_{\Phi_{38}}+\frac{24}{5} \eta_{\Phi_{39}}+\frac{24}{5} \eta_{\Phi_{40}}+\frac{2}{5} \eta_{\Phi_{41}}+\frac{49}{5} \eta_{\Phi_{42}}+\frac{12}{5} \eta_{\Phi_{43}}+\frac{64}{5} \eta_{\Phi_{44}}+\frac{24}{5} \eta_{\Phi_{45}}+\frac{3}{5} \eta_{\Phi_{46}}+\frac{4}{5} \eta_{\Phi_{47}}+\frac{1}{10} \eta_{\Phi_{48}}+\frac{3}{5} \eta_{\Phi_{49}}+\frac{4}{5} \eta_{\Phi_{50}}+\frac{1}{10} \eta_{\Phi_{51}} \, .
\end{dmath*}
\end{dgroup*}
Here, $PSV$ denotes the Pati-Salam gauge bosons in the $(\overline{3},1,-4)$ and $W_R$ the right-handed $W_R^\pm$ in the $(1,1,-6)$. 

\begin{table}[ph]
\centering
\begin{center}
\tiny
\begin{tabular}{|c|c|c|c|}
\hline
 $\bm{SO(10)}$ & $\bm{SU(5})$ & $\bm{SU(3)_C \times SU(2)_L \times U(1)_Y}$ & \textbf{Label} \\
\hline
\multirow{4}{*}{$10$} & \multirow{2}{*}{$5$} & $(1,2,3)$ & $\Phi_{1}$ \\
 &  & $(3,1,-2)$ & $\Phi_{2}$ \\
\hhline{~---}
 &  \multirow{2}{*}{$\overline{5}$} & $(1,2,-3)$ & H \\
 &  & $(\overline{3},1,2)$ & $\Phi_{3}$ \\
\hline
\end{tabular}  
\end{center}
\caption{Decomposition of the scalar $10$ representation of $SO(10)$ with respect to the subgroups $SU(5)$ and $SU(3)_C \times SU(2)_L \times U(1)_Y$. For further details, see Table~\ref{table:su5decomp}.}
\end{table}

\begin{table}[ph]
\centering
\begin{center}
\tiny
\begin{tabular}{|c|c|c|c|}
\hline
 $\bm{SO(10)}$ & $\bm{SU(5})$ & $\bm{SU(3)_C \times SU(2)_L \times U(1)_Y}$ & \textbf{Label} \\
\hline
\multirow{12}{*}{$45$} & 1 & $(1,1,0)$ & $s_{1}$ \\
\hhline{~---}
 & \multirow{3}{*}{$10$} & $(1,1,6)$ & $\Phi_{4}$ \\
 &  & $(\overline{3},1,-4)$ & $\Phi_{5}$ \\
 &  & $(3,2,1)$ & $\Phi_{6}$ \\
 \hhline{~---}
 & \multirow{3}{*}{$\overline{10}$} & $(1,1,-6)$ & $\Phi_{7}$ \\
 &  & $(3,1,4)$ & $\Phi_{8}$ \\
 &  & $(\overline{3},2,-1)$ & $\Phi_{9}$ \\
 \hhline{~---}
 & \multirow{5}{*}{$24$} & $(1,1,0)$ & $s_{2}$ \\
 &  & $(1,3,0)$ & $\Phi_{10}$ \\
 &  & $(3,2,-5)$ & $\xi_{1}$ \\
 &  & $(\overline{3},2,5)$ & $\xi_{2}$ \\
 &  & $(8,1,0)$ & $\Phi_{11}$ \\
\hline
\end{tabular}  
\end{center}
\caption{Decomposition of the scalar $45$ representation of $SO(10)$ with respect to the subgroups $SU(5)$ and $SU(3)_C \times SU(2)_L \times U(1)_Y$. For further details, see Table~\ref{table:su5decomp}.}
\end{table}

\begin{table}[ph]
\centering
\begin{center}
\tiny
\begin{tabular}{|c|c|c|c|}
\hline
 $\bm{SO(10)}$ & $\bm{SU(5})$ & $\bm{SU(3)_C \times SU(2)_L \times U(1)_Y}$ & \textbf{Label} \\
\hline
\multirow{24}{*}{$120$} & \multirow{2}{*}{$5$} & $(1,2,3)$ & $\Phi_{12}$ \\
 &  & $(3,1,-2)$ & $\Phi_{13}$ \\
\hhline{~---}
 & \multirow{2}{*}{$\overline{5}$} & $(1,2,-3)$ & $\Phi_{14}$ \\
 &  & $(\overline{3},1,2)$ & $\Phi_{15}$ \\
\hhline{~---}
 & \multirow{3}{*}{$10$} & $(1,1,6)$ & $\Phi_{16}$ \\
 &  & $(\overline{3},1,-4)$ & $\Phi_{17}$ \\
 &  & $(3,2,1)$ & $\Phi_{18}$ \\
\hhline{~---} 
 & \multirow{3}{*}{$\overline{10}$} & $(1,1,-6)$ & $\Phi_{19}$ \\
 &  & $(3,1,4)$ & $\Phi_{20}$ \\
 &  & $(\overline{3},2,-1)$ & $\Phi_{21}$ \\
\hhline{~---} 
 & \multirow{7}{*}{$45$} & $(1,2,3)$ & $\Phi_{22}$ \\
 &  & $(3,1,-2)$ & $\Phi_{23}$ \\
 &  & $(3,3,-2)$ & $\Phi_{24}$ \\
 &  & $(\overline{3},1,8)$ & $\Phi_{25}$ \\
 &  & $(\overline{3},2,-7)$ & $\Phi_{26}$ \\
 &  & $(\overline{6},1,-2)$ & $\Phi_{27}$ \\
 &  & $(8,2,3)$ & $\Phi_{28}$ \\
\hhline{~---} 
 & \multirow{7}{*}{$\overline{45}$} & $(1,2,-3)$ & $\Phi_{29}$ \\
 &  & $(\overline{3},1,2)$ & $\Phi_{30}$ \\
 &  & $(\overline{3},3,2)$ & $\Phi_{31}$ \\
 &  & $(3,1,-8)$ & $\Phi_{32}$ \\
 &  & $(3,2,7)$ & $\Phi_{33}$ \\
 &  & $(6,1,2)$ & $\Phi_{34}$ \\
 &  & $(8,2,-3)$ & $\Phi_{35}$ \\
\hline
\end{tabular}  
\end{center}
\caption{Decomposition of the scalar $120$ representation of $SO(10)$ with respect to the subgroups $SU(5)$ and $SU(3)_C \times SU(2)_L \times U(1)_Y$. For further details, see Table~\ref{table:su5decomp}.}
\end{table}

\begin{table}[ph]
\centering
\begin{center}
\small
\begin{tabular}{|c|c|c|c|}
\hline
 $\bm{SO(10)}$ & $\bm{SU(5})$ & $\bm{SU(3)_C \times SU(2)_L \times U(1)_Y}$ & \textbf{Label} \\
\hline
\multirow{21}{*}{$\overline{126}$} & 1 & $(1,1,0)$ & $s_{3}$ \\
 \hhline{~---} 
 & \multirow{2}{*}{$5$} & $(1,2,3)$ & $\Phi_{36}$ \\
 &  & $(3,1,-2)$ & $\Phi_{37}$ \\
 \hhline{~---} 
 & \multirow{3}{*}{$\overline{10}$} & $(1,1,-6)$ & $\xi_{3}$ \\
 &  & $(3,1,4)$ & $\xi_{4}$ \\
 &  & $(\overline{3},2,-1)$ & $\xi_{5}$ \\
 \hhline{~---} 
 & \multirow{3}{*}{$15$} & $(1,3,6)$ & $\Phi_{38}$ \\
 &  & $(3,2,1)$ & $\Phi_{39}$ \\
 &  & $(6,1,-4)$ & $\Phi_{40}$ \\
 \hhline{~---} 
 & \multirow{7}{*}{$\overline{45}$} & $(1,2,-3)$ & $\Phi_{41}$ \\
 &  & $(\overline{3},1,2)$ & $\Phi_{42}$ \\
 &  & $(\overline{3},3,2)$ & $\Phi_{43}$ \\
 &  & $(3,1,-8)$ & $\Phi_{44}$ \\
 &  & $(3,2,7)$ & $\Phi_{45}$ \\
 &  & $(6,1,2)$ & $\Phi_{46}$ \\
 &  & $(8,2,-3)$ & $\Phi_{47}$ \\
 \hhline{~---} 
 & \multirow{6}{*}{$50$} & $(1,1,-12)$ & $\Phi_{48}$ \\
 &  & $(3,1,-2)$ & $\Phi_{49}$ \\
 &  & $(\overline{3},2,-7)$ & $\Phi_{50}$ \\
 &  & $(\overline{6},3,-2)$ & $\Phi_{51}$ \\
 &  & $(6,1,8)$ & $\Phi_{52}$ \\
 &  & $(8,2,3)$ & $\Phi_{53}$ \\
\hline
\end{tabular}  
\end{center}
\caption{Decomposition of the scalar $\overline{126}$ representation of $SO(10)$ with respect to the subgroups $SU(5)$ and $SU(3)_C \times SU(2)_L \times U(1)_Y$. For further details, see Table~\ref{table:su5decomp}.}
\end{table}

\clearpage

\subsection{$E_6$}
  \label{app:e6thresholdsanddecomposition}

Using Eq.~\eqref{eq:thresholdformula}, we find for the threshold corrections in conservative $E_6$ GUTs

\begin{dgroup*}
\begin{dmath*}
\lambda_{3C} = 9 -21 \left( \eta_{PSV}+ \eta_{E_{2}}+ \eta_{E_{4}}\right)+\eta_{\Sigma_{2}}+\eta_{\Sigma_{3}}+\eta_{\Sigma_{5}}+\eta_{\Sigma_{7}}+2 \eta_{\Sigma_{8}}+\eta_{\Sigma_{10}}+\eta_{\Sigma_{12}}+\eta_{\Sigma_{14}}+\eta_{\Sigma_{16}}+2 \eta_{\Sigma_{17}}+\eta_{\Sigma_{19}}+\eta_{\Sigma_{21}}+2 \eta_{\Sigma_{22}}+\eta_{\Sigma_{24}}+2 \eta_{\Sigma_{25}}+\eta_{\Sigma_{27}}+2 \eta_{\Sigma_{28}}+6 \eta_{\Sigma_{30}}+\eta_{\Sigma_{32}}+\eta_{\Sigma_{34}}+\eta_{\Sigma_{36}}+2 \eta_{\Sigma_{37}}+\eta_{\Sigma_{39}}+2 \eta_{\Sigma_{40}}+\eta_{\Sigma_{42}}+3 \eta_{\Sigma_{43}}+\eta_{\Sigma_{44}}+2 \eta_{\Sigma_{45}}+5 \eta_{\Sigma_{46}}+12 \eta_{\Sigma_{47}}+\eta_{\Sigma_{49}}+3 \eta_{\Sigma_{50}}+\eta_{\Sigma_{51}}+2 \eta_{\Sigma_{52}}+5 \eta_{\Sigma_{53}}+12 \eta_{\Sigma_{54}}+\eta_{\Sigma_{56}}+\eta_{\Sigma_{58}}+\eta_{\Sigma_{60}}+2 \eta_{\Sigma_{61}}+2 \eta_{\Sigma_{63}}+5 \eta_{\Sigma_{64}}+2 \eta_{\Sigma_{66}}+2 \eta_{\Sigma_{67}}+6 \eta_{\Sigma_{68}}+2 \eta_{\Sigma_{70}}+\eta_{\Sigma_{71}}+3 \eta_{\Sigma_{72}}+6 \eta_{\Sigma_{73}}+10 \eta_{\Sigma_{74}}+\eta_{\Sigma_{76}}+3 \eta_{\Sigma_{77}}+\eta_{\Sigma_{78}}+2 \eta_{\Sigma_{79}}+5 \eta_{\Sigma_{80}}+12 \eta_{\Sigma_{81}}+\eta_{\Sigma_{83}}+\eta_{\Sigma_{85}}+2 \eta_{\Sigma_{87}}+5 \eta_{\Sigma_{88}}+2 \eta_{\Sigma_{90}}+5 \eta_{\Sigma_{91}}+2 \eta_{\Sigma_{93}}+2 \eta_{\Sigma_{94}}+6 \eta_{\Sigma_{95}}+\eta_{\Sigma_{97}}+2 \eta_{\Sigma_{99}}+5 \eta_{\Sigma_{100}}+\eta_{\Sigma_{102}}+3 \eta_{\Sigma_{103}}+\eta_{\Sigma_{104}}+2 \eta_{\Sigma_{105}}+5 \eta_{\Sigma_{106}}+12 \eta_{\Sigma_{107}}+\eta_{\Sigma_{109}}+2 \eta_{\Sigma_{110}}+15 \eta_{\Sigma_{111}}+5 \eta_{\Sigma_{112}}+12 \eta_{\Sigma_{113}}+\eta_{\Sigma_{115}}+\eta_{\Sigma_{117}}+\eta_{\Sigma_{119}}+2 \eta_{\Sigma_{120}}+2 \eta_{\Sigma_{122}}+5 \eta_{\Sigma_{123}}+2 \eta_{\Sigma_{125}}+2 \eta_{\Sigma_{126}}+6 \eta_{\Sigma_{127}}+2 \eta_{\Sigma_{129}}+\eta_{\Sigma_{130}}+3 \eta_{\Sigma_{131}}+6 \eta_{\Sigma_{132}}+10 \eta_{\Sigma_{133}}+\eta_{\Sigma_{135}}+3 \eta_{\Sigma_{136}}+\eta_{\Sigma_{137}}+2 \eta_{\Sigma_{138}}+5 \eta_{\Sigma_{139}}+12 \eta_{\Sigma_{140}} + 8 \left( \eta_{D_{1}}+ \eta_{D_{2}}+ \eta_{D_{3}}\right)\, , \end{dmath*}
\begin{dmath*}
 \lambda_{2L}  = 10 -21  \eta_{E_{1}}+\eta_{\Sigma_{1}}+\eta_{\Sigma_{4}}+3 \eta_{\Sigma_{8}}+\eta_{\Sigma_{9}}+\eta_{\Sigma_{11}}+\eta_{\Sigma_{13}}+3 \eta_{\Sigma_{17}}+\eta_{\Sigma_{18}}+3 \eta_{\Sigma_{22}}+3 \eta_{\Sigma_{25}}+3 \eta_{\Sigma_{28}}+4 \eta_{\Sigma_{29}}+\eta_{\Sigma_{31}}+\eta_{\Sigma_{33}}+3 \eta_{\Sigma_{37}}+3 \eta_{\Sigma_{40}}+\eta_{\Sigma_{41}}+12 \eta_{\Sigma_{43}}+3 \eta_{\Sigma_{45}}+8 \eta_{\Sigma_{47}}+\eta_{\Sigma_{48}}+12 \eta_{\Sigma_{50}}+3 \eta_{\Sigma_{52}}+8 \eta_{\Sigma_{54}}+\eta_{\Sigma_{55}}+\eta_{\Sigma_{57}}+3 \eta_{\Sigma_{61}}+4 \eta_{\Sigma_{62}}+3 \eta_{\Sigma_{63}}+4 \eta_{\Sigma_{65}}+3 \eta_{\Sigma_{66}}+3 \eta_{\Sigma_{67}}+\eta_{\Sigma_{69}}+3 \eta_{\Sigma_{70}}+12 \eta_{\Sigma_{72}}+6 \eta_{\Sigma_{74}}+\eta_{\Sigma_{75}}+12 \eta_{\Sigma_{77}}+3 \eta_{\Sigma_{79}}+8 \eta_{\Sigma_{81}}+\eta_{\Sigma_{82}}+\eta_{\Sigma_{84}}+4 \eta_{\Sigma_{86}}+3 \eta_{\Sigma_{87}}+4 \eta_{\Sigma_{89}}+3 \eta_{\Sigma_{90}}+4 \eta_{\Sigma_{92}}+3 \eta_{\Sigma_{93}}+3 \eta_{\Sigma_{94}}+\eta_{\Sigma_{96}}+4 \eta_{\Sigma_{98}}+3 \eta_{\Sigma_{99}}+\eta_{\Sigma_{101}}+12 \eta_{\Sigma_{103}}+3 \eta_{\Sigma_{105}}+8 \eta_{\Sigma_{107}}+3 \eta_{\Sigma_{110}}+24 \eta_{\Sigma_{111}}+8 \eta_{\Sigma_{113}}+\eta_{\Sigma_{114}}+\eta_{\Sigma_{116}}+3 \eta_{\Sigma_{120}}+4 \eta_{\Sigma_{121}}+3 \eta_{\Sigma_{122}}+4 \eta_{\Sigma_{124}}+3 \eta_{\Sigma_{125}}+3 \eta_{\Sigma_{126}}+\eta_{\Sigma_{128}}+3 \eta_{\Sigma_{129}}+12 \eta_{\Sigma_{131}}+6 \eta_{\Sigma_{133}}+\eta_{\Sigma_{134}}+12 \eta_{\Sigma_{136}}+3 \eta_{\Sigma_{138}}+8 \eta_{\Sigma_{140}} + 8 \left( \eta_{L_{1}}+\eta_{L_{2}}+\eta_{L_{3}}\right)\, ,  
 \end{dmath*}
 \begin{dmath*}
  \lambda_{1Y}  = 12 -21 \left(\frac{6}{5} \eta_{W_{R}}+\frac{8}{5} \eta_{PSV}+\frac{3}{5} \eta_{E_{1}}+\frac{2}{5} \eta_{E_{2}}+\frac{6}{5} \eta_{E_{3}}+\frac{8}{5} \eta_{E_{4}}\right)+ \frac{3}{5} \eta_{\Sigma_{1}}+\frac{2}{5} \eta_{\Sigma_{2}}+\frac{2}{5} \eta_{\Sigma_{3}}+\frac{3}{5} \eta_{\Sigma_{4}}+\frac{2}{5} \eta_{\Sigma_{5}}+\frac{6}{5} \eta_{\Sigma_{6}}+\frac{8}{5} \eta_{\Sigma_{7}}+\frac{1}{5} \eta_{\Sigma_{8}}+\frac{3}{5} \eta_{\Sigma_{9}}+\frac{2}{5} \eta_{\Sigma_{10}}+\frac{3}{5} \eta_{\Sigma_{11}}+\frac{2}{5} \eta_{\Sigma_{12}}+\frac{3}{5} \eta_{\Sigma_{13}}+\frac{2}{5} \eta_{\Sigma_{14}}+\frac{6}{5} \eta_{\Sigma_{15}}+\frac{8}{5} \eta_{\Sigma_{16}}+\frac{1}{5} \eta_{\Sigma_{17}}+\frac{3}{5} \eta_{\Sigma_{18}}+\frac{2}{5} \eta_{\Sigma_{19}}+\frac{6}{5} \eta_{\Sigma_{20}}+\frac{8}{5} \eta_{\Sigma_{21}}+\frac{1}{5} \eta_{\Sigma_{22}}+\frac{6}{5} \eta_{\Sigma_{23}}+\frac{8}{5} \eta_{\Sigma_{24}}+\frac{1}{5} \eta_{\Sigma_{25}}+\frac{6}{5} \eta_{\Sigma_{26}}+\frac{8}{5} \eta_{\Sigma_{27}}+\frac{1}{5} \eta_{\Sigma_{28}}+\frac{3}{5} \eta_{\Sigma_{31}}+\frac{2}{5} \eta_{\Sigma_{32}}+\frac{3}{5} \eta_{\Sigma_{33}}+\frac{2}{5} \eta_{\Sigma_{34}}+\frac{6}{5} \eta_{\Sigma_{35}}+\frac{8}{5} \eta_{\Sigma_{36}}+\frac{1}{5} \eta_{\Sigma_{37}}+\frac{6}{5} \eta_{\Sigma_{38}}+\frac{8}{5} \eta_{\Sigma_{39}}+\frac{1}{5} \eta_{\Sigma_{40}}+\frac{3}{5} \eta_{\Sigma_{41}}+\frac{2}{5} \eta_{\Sigma_{42}}+\frac{6}{5} \eta_{\Sigma_{43}}+\frac{32}{5} \eta_{\Sigma_{44}}+\frac{49}{5} \eta_{\Sigma_{45}}+\frac{4}{5} \eta_{\Sigma_{46}}+\frac{24}{5} \eta_{\Sigma_{47}}+\frac{3}{5} \eta_{\Sigma_{48}}+\frac{2}{5} \eta_{\Sigma_{49}}+\frac{6}{5} \eta_{\Sigma_{50}}+\frac{32}{5} \eta_{\Sigma_{51}}+\frac{49}{5} \eta_{\Sigma_{52}}+\frac{4}{5} \eta_{\Sigma_{53}}+\frac{24}{5} \eta_{\Sigma_{54}}+\frac{3}{5} \eta_{\Sigma_{55}}+\frac{2}{5} \eta_{\Sigma_{56}}+\frac{3}{5} \eta_{\Sigma_{57}}+\frac{2}{5} \eta_{\Sigma_{58}}+\frac{6}{5} \eta_{\Sigma_{59}}+\frac{8}{5} \eta_{\Sigma_{60}}+\frac{1}{5} \eta_{\Sigma_{61}}+\frac{18}{5} \eta_{\Sigma_{62}}+\frac{1}{5} \eta_{\Sigma_{63}}+\frac{16}{5} \eta_{\Sigma_{64}}+5 \eta_{\Sigma_{66}}+5 \eta_{\Sigma_{67}}+\frac{27}{5} \eta_{\Sigma_{69}}+\frac{1}{5} \eta_{\Sigma_{70}}+\frac{8}{5} \eta_{\Sigma_{71}}+\frac{24}{5} \eta_{\Sigma_{72}}+\frac{48}{5} \eta_{\Sigma_{73}}+\frac{2}{5} \eta_{\Sigma_{74}}+\frac{3}{5} \eta_{\Sigma_{75}}+\frac{2}{5} \eta_{\Sigma_{76}}+\frac{6}{5} \eta_{\Sigma_{77}}+\frac{32}{5} \eta_{\Sigma_{78}}+\frac{49}{5} \eta_{\Sigma_{79}}+\frac{4}{5} \eta_{\Sigma_{80}}+\frac{24}{5} \eta_{\Sigma_{81}}+\frac{3}{5} \eta_{\Sigma_{82}}+\frac{2}{5} \eta_{\Sigma_{83}}+\frac{3}{5} \eta_{\Sigma_{84}}+\frac{2}{5} \eta_{\Sigma_{85}}+\frac{18}{5} \eta_{\Sigma_{86}}+\frac{1}{5} \eta_{\Sigma_{87}}+\frac{16}{5} \eta_{\Sigma_{88}}+\frac{18}{5} \eta_{\Sigma_{89}}+\frac{1}{5} \eta_{\Sigma_{90}}+\frac{16}{5} \eta_{\Sigma_{91}}+5 \eta_{\Sigma_{93}}+5 \eta_{\Sigma_{94}}+\frac{3}{5} \eta_{\Sigma_{96}}+\frac{2}{5} \eta_{\Sigma_{97}}+\frac{18}{5} \eta_{\Sigma_{98}}+\frac{1}{5} \eta_{\Sigma_{99}}+\frac{16}{5} \eta_{\Sigma_{100}}+\frac{3}{5} \eta_{\Sigma_{101}}+\frac{2}{5} \eta_{\Sigma_{102}}+\frac{6}{5} \eta_{\Sigma_{103}}+\frac{32}{5} \eta_{\Sigma_{104}}+\frac{49}{5} \eta_{\Sigma_{105}}+\frac{4}{5} \eta_{\Sigma_{106}}+\frac{24}{5} \eta_{\Sigma_{107}}+\frac{24}{5} \eta_{\Sigma_{108}}+\frac{2}{5} \eta_{\Sigma_{109}}+\frac{49}{5} \eta_{\Sigma_{110}}+\frac{12}{5} \eta_{\Sigma_{111}}+\frac{64}{5} \eta_{\Sigma_{112}}+\frac{24}{5} \eta_{\Sigma_{113}}+\frac{3}{5} \eta_{\Sigma_{114}}+\frac{2}{5} \eta_{\Sigma_{115}}+\frac{3}{5} \eta_{\Sigma_{116}}+\frac{2}{5} \eta_{\Sigma_{117}}+\frac{6}{5} \eta_{\Sigma_{118}}+\frac{8}{5} \eta_{\Sigma_{119}}+\frac{1}{5} \eta_{\Sigma_{120}}+\frac{18}{5} \eta_{\Sigma_{121}}+\frac{1}{5} \eta_{\Sigma_{122}}+\frac{16}{5} \eta_{\Sigma_{123}}+5 \eta_{\Sigma_{125}}+5 \eta_{\Sigma_{126}}+\frac{27}{5} \eta_{\Sigma_{128}}+\frac{1}{5} \eta_{\Sigma_{129}}+\frac{8}{5} \eta_{\Sigma_{130}}+\frac{24}{5} \eta_{\Sigma_{131}}+\frac{48}{5} \eta_{\Sigma_{132}}+\frac{2}{5} \eta_{\Sigma_{133}}+\frac{3}{5} \eta_{\Sigma_{134}}+\frac{2}{5} \eta_{\Sigma_{135}}+\frac{6}{5} \eta_{\Sigma_{136}}+\frac{32}{5} \eta_{\Sigma_{137}}+\frac{49}{5} \eta_{\Sigma_{138}}+\frac{4}{5} \eta_{\Sigma_{139}}+\frac{24}{5} \eta_{\Sigma_{140}} +8 \left(\frac{2}{5} \eta_{D_{1}}+\frac{3}{5} \eta_{L_1}  +\frac{2}{5} \eta_{D_{2}}+\frac{3}{5} \eta_{L_2}+\frac{2}{5} \eta_{D_{3}}+\frac{3}{5} \eta_{L_3}  \right) \, .
\end{dmath*}
\end{dgroup*}

Here, $PSV$ denotes the Pati-Salam gauge bosons in the $(\overline{3},1,-4)$, $W_R$ the right-handed $W_R^\pm$ in the $(1,1,-6)$. In addition, $E_i$ are the additional $E_6$ gauge bosons in the $(1,2,-3)$, $(\overline{3},1,2)$, $(1,1,6)$, $(\overline{3},1,-4)$ respectively. $D_i$ and $L_i$ denote the three generations of vector-like quarks and leptons.


\begin{table}[ph]
\centering
\begin{center}
\small
\begin{tabular}{|c|c|c|c|c|}
\hline
$\bm{E_6}$ & $\bm{SO(10)}$ & $\bm{SU(5})$ & $\bm{SU(3)_C \times SU(2)_L \times U(1)_Y}$ & \textbf{Label} \\
\hline
\multirow{11}{*}{$27$} & $1$ & $1$ & $(1,1,0)$ & $s_{1}$ \\
\hhline{~----}
& \multirow{4}{*}{$10$} & \multirow{2}{*}{$5$} & $(1,2,3)$ & $\Sigma_{1}$ \\
& &  & $(3,1,-2)$ & $\Sigma_{2}$ \\
\hhline{~~---}
& & \multirow{2}{*}{$\overline{5}$} & $(1,2,-3)$ & H \\
& &  & $(\overline{3},1,2)$ & $\Sigma_{3}$ \\
\hhline{~----}
& \multirow{6}{*}{$16$} & 1 & $(1,1,0)$ & $s_{2}$ \\
\hhline{~~---}
& & \multirow{2}{*}{$\overline{5}$} & $(1,2,-3)$ & $\Sigma_{4}$ \\
& &  & $(\overline{3},1,2)$ & $\Sigma_{5}$ \\
\hhline{~~---}
& & \multirow{3}{*}{$10$} & $(1,1,6)$ & $\Sigma_{6}$ \\
& &  & $(\overline{3},1,-4)$ & $\Sigma_{7}$ \\
& &  & $(3,2,1)$ & $\Sigma_{8}$ \\
\hline
\end{tabular}  
\end{center}
\caption{Decomposition of the scalar $27$-dimensional representation of $E_6$ with respect to the subgroups $SO(10)$, $SU(5)$ and $SU(3)_C \times SU(2)_L \times U(1)_Y$. For further details, see Table~\ref{table:su5decomp}.}
\end{table}

\begin{table}[ph]
\centering
\begin{center}
\begin{tabular}{|c|c|c|c|c|c|c|}
\hline
$\bm{E_6}$ & $\bm{SO(10)}$ &$\bm{SU(5})$ & $\bm{SU(3)_C \times SU(2)_L \times U(1)_Y}$ & $\bm{A_{23}/r_I}$ & $\bm{A_{12}/r_I}$  \\
\hline
\multirow{5}{*}{$27$} & 1 & 1 & $(1,1,0)$ & $0$ & $0$ \\
 \hhline{~-----}
 & \multirow{4}{*}{$10$} & \multirow{2}{*}{$5$} & $(1,2,3)$ & $1/3$ & $-2/15$ \\
 \hhline{~~~---}
 &  &  & $(3,1,-2)$ & $-1/3$ & $2/15$ \\
  \hhline{~~----}
 &  &\multirow{2}{*}{$\overline{5}$} & $(1,2,-3)$ & $1/3$ & $-2/15$ \\
 \hhline{~~~---}
 &  &  & $(\overline{3},1,2)$ & $-1/3$ & $2/15$ \\
 \hline
\end{tabular}
\end{center}
\caption{Contributions of the exotic fermions in the fundamental $27$-dimensional representation of $E_6$ to the ratio $A_{23}/A_{12}$. }
\label{table:e6fermioncontributions}
\end{table}

\renewcommand\arraystretch{1.1}
\begin{table}[ph]
\centering
\begin{center}
\tiny
\begin{tabular}{|c|c|c|c|c|}
\hline
$\bm{E_6}$ & $\bm{SO(10)}$ & $\bm{SU(5})$ & $\bm{SU(3)_C \times SU(2)_L \times U(1)_Y}$ & \textbf{Label} \\
\hline
\multirow{79}{*}{$351$} & \multirow{4}{*}{$10$} & \multirow{2}{*}{$5$} & $(1,2,3)$ & $\Sigma_{9}$ \\
& &  & $(3,1,-2)$ & $\Sigma_{10}$ \\
& &\multirow{2}{*}{$\overline{5}$} & $(1,2,-3)$ & $\Sigma_{11}$ \\
& &  & $(\overline{3},1,2)$ & $\Sigma_{12}$ \\
\hhline{~----}
& \multirow{6}{*}{$16$} & 1 & $(1,1,0)$ & $s_{3}$ \\
\hhline{~~---}
& & \multirow{2}{*}{$\overline{5}$}& $(1,2,-3)$ & $\Sigma_{13}$ \\
& &  & $(\overline{3},1,2)$ & $\Sigma_{14}$ \\
\hhline{~~---}
& & \multirow{3}{*}{$10$} & $(1,1,6)$ & $\Sigma_{15}$ \\
& &  & $(\overline{3},1,-4)$ & $\Sigma_{16}$ \\
& &  & $(3,2,1)$ & $\Sigma_{17}$ \\
\hhline{~----}
& \multirow{6}{*}{$\overline{16}$} & 1 & $(1,1,0)$ & $s_{4}$ \\
\hhline{~~---}
& & \multirow{2}{*}{$5$} & $(1,2,3)$ & $\Sigma_{18}$ \\
& &  & $(3,1,-2)$ & $\Sigma_{19}$ \\
\hhline{~~---}
& & \multirow{3}{*}{$\overline{10}$} & $(1,1,-6)$ & $\Sigma_{20}$ \\
& &  & $(3,1,4)$ & $\Sigma_{21}$ \\
& &  & $(\overline{3},2,-1)$ & $\Sigma_{22}$ \\
\hhline{~----}
& \multirow{12}{*}{$45$} & 1 & $(1,1,0)$ & $s_{5}$ \\
\hhline{~~---}
& & \multirow{3}{*}{$10$} & $(1,1,6)$ & $\Sigma_{23}$ \\
& &  & $(\overline{3},1,-4)$ & $\Sigma_{24}$ \\
& &  & $(3,2,1)$ & $\Sigma_{25}$ \\
\hhline{~~---}
& & \multirow{3}{*}{$\overline{10}$} & $(1,1,-6)$ & $\Sigma_{26}$ \\
& &  & $(3,1,4)$ & $\Sigma_{27}$ \\
& &  & $(\overline{3},2,-1)$ & $\Sigma_{28}$ \\
\hhline{~~---}
& & \multirow{5}{*}{$24$} & $(1,1,0)$ & $s_{6}$ \\
& &  & $(1,3,0)$ & $\Sigma_{29}$ \\
& &  & $(3,2,-5)$ & $\xi_{1}$ \\
& &  & $(\overline{3},2,5)$ & $\xi_{2}$ \\
& &  & $(8,1,0)$ & $\Sigma_{30}$ \\
\hhline{~----}
& \multirow{23}{*}{$120$} & \multirow{2}{*}{$5$} & $(1,2,3)$ & $\Sigma_{31}$ \\
& &  & $(3,1,-2)$ & $\Sigma_{32}$ \\
\hhline{~~---}
& & \multirow{2}{*}{$\overline{5}$}& $(1,2,-3)$ & $\Sigma_{33}$ \\
& &  & $(\overline{3},1,2)$ & $\Sigma_{34}$ \\
\hhline{~~---}
& & \multirow{3}{*}{$10$} & $(1,1,6)$ & $\Sigma_{35}$ \\
& &  & $(\overline{3},1,-4)$ & $\Sigma_{36}$ \\
& &  & $(3,2,1)$ & $\Sigma_{37}$ \\
\hhline{~~---}
& & \multirow{3}{*}{$\overline{10}$} & $(1,1,-6)$ & $\Sigma_{38}$ \\
& &  & $(3,1,4)$ & $\Sigma_{39}$ \\
& &  & $(\overline{3},2,-1)$ & $\Sigma_{40}$ \\
\hhline{~~---}
& & \multirow{7}{*}{$45$} & $(1,2,3)$ & $\Sigma_{41}$ \\
& & & $(3,1,-2)$ & $\Sigma_{42}$ \\
& & & $(3,3,-2)$ & $\Sigma_{43}$ \\
& & & $(\overline{3},1,8)$ & $\Sigma_{44}$ \\
& & & $(\overline{3},2,-7)$ & $\Sigma_{45}$ \\
& & & $(\overline{6},1,-2)$ & $\Sigma_{46}$ \\
& & & $(8,2,3)$ & $\Sigma_{47}$ \\
\hhline{~~---}
& & \multirow{7}{*}{$\overline{45}$} & $(1,2,-3)$ & $\Sigma_{48}$ \\
& & & $(\overline{3},1,2)$ & $\Sigma_{49}$ \\
& & & $(\overline{3},3,2)$ & $\Sigma_{50}$ \\
& & & $(3,1,-8)$ & $\Sigma_{51}$ \\
& & & $(3,2,7)$ & $\Sigma_{52}$ \\
& & & $(6,1,2)$ & $\Sigma_{53}$ \\
& & & $(8,2,-3)$ & $\Sigma_{54}$ \\
\hhline{~----}
& \multirow{27}{*}{$144$} & \multirow{2}{*}{$5$} & $(1,2,3)$ & $\Sigma_{55}$ \\
& &  & $(3,1,-2)$ & $\Sigma_{56}$ \\
\hhline{~~---}
& & \multirow{2}{*}{$\overline{5}$} & $(1,2,-3)$ & $\Sigma_{57}$ \\
& &  & $(\overline{3},1,2)$ & $\Sigma_{58}$ \\
\hhline{~~---}
& & \multirow{3}{*}{$10$} & $(1,1,6)$ & $\Sigma_{59}$ \\
& & & $(\overline{3},1,-4)$ & $\Sigma_{60}$ \\
& & & $(3,2,1)$ & $\Sigma_{61}$ \\
\hhline{~~---}
& & \multirow{3}{*}{$15$} & $(1,3,6)$ & $\Sigma_{62}$ \\
& & & $(3,2,1)$ & $\Sigma_{63}$ \\
& & & $(6,1,-4)$ & $\Sigma_{64}$ \\
\hhline{~~---}
& & \multirow{5}{*}{$24$} & $(1,1,0)$ & $s_{7}$ \\
& & & $(1,3,0)$ & $\Sigma_{65}$ \\
& & & $(3,2,-5)$ & $\Sigma_{66}$ \\
& & & $(\overline{3},2,5)$ & $\Sigma_{67}$ \\
& & & $(8,1,0)$ & $\Sigma_{68}$ \\
\hhline{~~---}
& & \multirow{6}{*}{$40$} & $(1,2,-9)$ & $\Sigma_{69}$ \\
& & & $(3,2,1)$ & $\Sigma_{70}$ \\
& & & $(\overline{3},1,-4)$ & $\Sigma_{71}$ \\
& & & $(\overline{3},3,-4)$ & $\Sigma_{72}$ \\
& & & $(8,1,6)$ & $\Sigma_{73}$ \\
& & & $(\overline{6},2,1)$ & $\Sigma_{74}$ \\
\hhline{~~---}
& & \multirow{7}{*}{$\overline{45}$} & $(1,2,-3)$ & $\Sigma_{75}$ \\
& & & $(\overline{3},1,2)$ & $\Sigma_{76}$ \\
& & & $(\overline{3},3,2)$ & $\Sigma_{77}$ \\
& & & $(3,1,-8)$ & $\Sigma_{78}$ \\
& & & $(3,2,7)$ & $\Sigma_{79}$ \\
& & & $(6,1,2)$ & $\Sigma_{80}$ \\
& & & $(8,2,-3)$ & $\Sigma_{81}$ \\
\hline
\end{tabular}
\end{center}
\caption{Decomposition of the $351$ representation of $E_6$ with respect to the subgroups $SO(10)$, $SU(5)$ and $SU(3)_C \times SU(2)_L \times U(1)_Y$. For further details, see Table~\ref{table:su5decomp}.}
\end{table}

\begin{table}[ph]
\centering
\begin{center}
\tiny
\begin{tabular}{|c|c|c|c|c|}
\hline
$\bm{E_6}$ & $\bm{SO(10)}$ & $\bm{SU(5})$ & $\bm{SU(3)_C \times SU(2)_L \times U(1)_Y}$ & \textbf{Label} \\
\hline
\multirow{71}{*}{$351'$} & 1 & 1 & $(1,1,0)$ & $s_{8}$ \\
\hhline{~----}
& \multirow{4}{*}{$10$} & \multirow{2}{*}{$5$} & $(1,2,3)$ & $\Sigma_{82}$ \\
& &  & $(3,1,-2)$ & $\Sigma_{83}$ \\
\hhline{~~---}
& & \multirow{2}{*}{$\overline{5}$}& $(1,2,-3)$ & $\Sigma_{84}$ \\
& &  & $(\overline{3},1,2)$ & $\Sigma_{85}$ \\
\hhline{~----}
& \multirow{6}{*}{$\overline{16}$} & 1 & $(1,1,0)$ & $s_{9}$ \\
\hhline{~~---}
& & \multirow{2}{*}{$5$} & $(1,2,3)$ & $\xi_{3}$ \\
& &  & $(3,1,-2)$ & $\xi_{4}$ \\
\hhline{~~---}
& & \multirow{3}{*}{$\overline{10}$} & $(1,1,-6)$ & $\xi_{5}$ \\
& &  & $(3,1,4)$ & $\xi_{6}$ \\
& &  & $(\overline{3},2,-1)$ & $\xi_{7}$ \\
\hhline{~----}
& \multirow{11}{*}{$54$} & \multirow{3}{*}{$15$} & $(1,3,6)$ & $\Sigma_{86}$ \\
& &  & $(3,2,1)$ & $\Sigma_{87}$ \\
& &  & $(6,1,-4)$ & $\Sigma_{88}$ \\
\hhline{~~---}
& & \multirow{3}{*}{$\overline{15}$} & $(1,3,-6)$ & $\Sigma_{89}$ \\
& &  & $(\overline{3},2,-1)$ & $\Sigma_{90}$ \\
& & & $(\overline{6},1,4)$ & $\Sigma_{91}$ \\
\hhline{~~---}
& &  \multirow{5}{*}{$\overline{24}$} & $(1,1,0)$ & $s_{10}$ \\
& & & $(1,3,0)$ & $\Sigma_{92}$ \\
& & & $(3,2,-5)$ & $\Sigma_{93}$ \\
& & & $(\overline{3},2,5)$ & $\Sigma_{94}$ \\
& & & $(8,1,0)$ & $\Sigma_{95}$ \\
\hhline{~----}
& \multirow{21}{*}{$\overline{126}$}  & 1 & $(1,1,0)$ & $s_{11}$ \\
\hhline{~~---}
& & \multirow{2}{*}{$5$} & $(1,2,3)$ & $\Sigma_{96}$ \\
& &  & $(3,1,-2)$ & $\Sigma_{97}$ \\
\hhline{~~---}
& & \multirow{3}{*}{$\overline{10}$} & $(1,1,-6)$ & $\xi_{8}$ \\
& &  & $(3,1,4)$ & $\xi_{9}$ \\
& &  & $(\overline{3},2,-1)$ & $\xi_{10}$ \\
\hhline{~~---}
& & \multirow{3}{*}{$15$} & $(1,3,6)$ & $\Sigma_{98}$ \\
& &  & $(3,2,1)$ & $\Sigma_{99}$ \\
& &  & $(6,1,-4)$ & $\Sigma_{100}$ \\
\hhline{~~---}
& & \multirow{7}{*}{$45$} & $(1,2,-3)$ & $\Sigma_{101}$ \\
& & & $(\overline{3},1,2)$ & $\Sigma_{102}$ \\
& & & $(\overline{3},3,2)$ & $\Sigma_{103}$ \\
& & & $(3,1,-8)$ & $\Sigma_{104}$ \\
& & & $(3,2,7)$ & $\Sigma_{105}$ \\
& & & $(6,1,2)$ & $\Sigma_{106}$ \\
& & & $(8,2,-3)$ & $\Sigma_{107}$ \\
\hhline{~~---}
& & \multirow{6}{*}{$50$} & $(1,1,-12)$ & $\Sigma_{108}$ \\
& & & $(3,1,-2)$ & $\Sigma_{109}$ \\
& & & $(\overline{3},2,-7)$ & $\Sigma_{110}$ \\
& & & $(\overline{6},3,-2)$ & $\Sigma_{111}$ \\
& & & $(6,1,8)$ & $\Sigma_{112}$ \\
& & & $(8,2,3)$ & $\Sigma_{113}$ \\
\hhline{~----}
& \multirow{27}{*}{$144$} &\multirow{2}{*}{$5$} & $(1,2,3)$ & $\Sigma_{114}$ \\
& &  & $(3,1,-2)$ & $\Sigma_{115}$ \\
\hhline{~~---}
& & \multirow{2}{*}{$\overline{5}$} & $(1,2,-3)$ & $\Sigma_{116}$ \\
& &  & $(\overline{3},1,2)$ & $\Sigma_{117}$ \\
\hhline{~~---}
& & \multirow{3}{*}{$10$} & $(1,1,6)$ & $\Sigma_{118}$ \\
& &  & $(\overline{3},1,-4)$ & $\Sigma_{119}$ \\
& &  & $(3,2,1)$ & $\Sigma_{120}$ \\
\hhline{~~---}
& & \multirow{3}{*}{$15$} & $(1,3,6)$ & $\Sigma_{121}$ \\
& &  & $(3,2,1)$ & $\Sigma_{122}$ \\
& &  & $(6,1,-4)$ & $\Sigma_{123}$ \\
\hhline{~~---}
& & \multirow{5}{*}{$24$} & $(1,1,0)$ & $s_{12}$ \\
& & & $(1,3,0)$ & $\Sigma_{124}$ \\
& & & $(3,2,-5)$ & $\Sigma_{125}$ \\
& & & $(\overline{3},2,5)$ & $\Sigma_{126}$ \\
& & & $(8,1,0)$ & $\Sigma_{127}$ \\
\hhline{~~---}
& & \multirow{6}{*}{$40$} & $(1,2,-9)$ & $\Sigma_{128}$ \\
& & & $(3,2,1)$ & $\Sigma_{129}$ \\
& & & $(\overline{3},1,-4)$ & $\Sigma_{130}$ \\
& & & $(\overline{3},3,-4)$ & $\Sigma_{131}$ \\
& & & $(8,1,6)$ & $\Sigma_{132}$ \\
& & & $(\overline{6},2,1)$ & $\Sigma_{133}$ \\
\hhline{~~---}
& & \multirow{7}{*}{$\overline{45}$}  & $(1,2,-3)$ & $\Sigma_{134}$ \\
& & & $(\overline{3},1,2)$ & $\Sigma_{135}$ \\
& & & $(\overline{3},3,2)$ & $\Sigma_{136}$ \\
& & & $(3,1,-8)$ & $\Sigma_{137}$ \\
& & & $(3,2,7)$ & $\Sigma_{138}$ \\
& & & $(6,1,2)$ & $\Sigma_{139}$ \\
& & & $(8,2,-3)$ & $\Sigma_{140}$ \\
\hline
\end{tabular}  
\end{center}
\caption{Decomposition of the $351'$ representation of $E_6$ with respect to the subgroups $SO(10)$, $SU(5)$ and $SU(3)_C \times SU(2)_L \times U(1)_Y$. For further details, see Table~\ref{table:su5decomp}.}
\end{table}

\clearpage

\subsection{$SO(10) \to SU(4)_C \times SU(2)_L \times U(1)_R$}

\label{app:so10421thresholdsanddecomposition}

Using Eq.~\eqref{eq:thresholdformula}, we find for the threshold corrections at the $SO(10)$ scale

\begin{dgroup*}
\begin{dmath*}
\lambda_{4C} = 4+ 2 \eta_{\zeta_{1}}+8 \eta_{\zeta_{4}}+6 \eta_{\zeta_{7}}+6 \eta_{\zeta_{8}}+2 \eta_{\zeta_{9}}+2 \eta_{\zeta_{10}}+2 \eta_{\zeta_{11}}+6 \eta_{\zeta_{12}}+16 \eta_{\zeta_{13}}+16 \eta_{\zeta_{14}}+2 \eta_{\zeta_{15}}+18 \eta_{\zeta_{16}}+6 \eta_{\zeta_{17}}+6 \eta_{\zeta_{18}}+16 \eta_{\zeta_{19}} \, , 
\end{dmath*}
\begin{dmath*}
 \lambda_{2L}  = 6+ \eta_{\zeta_{2}}+4 \eta_{\zeta_{3}}+\eta_{\zeta_{5}}+\eta_{\zeta_{6}}+24 \eta_{\zeta_{12}}+15 \eta_{\zeta_{13}}+15 \eta_{\zeta_{14}}+40 \eta_{\zeta_{16}}+15 \eta_{\zeta_{19}} \, ,  
 \end{dmath*}
 \begin{dmath*}
  \lambda_{1R}  =8+\eta_{\zeta_{2}}+\eta_{\zeta_{5}}+\eta_{\zeta_{6}}+12 \eta_{\zeta_{9}}+12 \eta_{\zeta_{11}}+36 \eta_{\zeta_{12}}+15 \eta_{\zeta_{13}}+15 \eta_{\zeta_{14}}+20 \eta_{\zeta_{17}}+15 \eta_{\zeta_{19}} \, 
\end{dmath*}
\end{dgroup*}

and for the corrections at the $SU(4)_C \times SU(2)_L  \times U(1)_R $ scale

\begin{dgroup*}
\begin{dmath*}
\lambda_{3C} = 1 -21 \left( \eta_{PSV} \right) +2 \eta_{\zeta_{1}}+2 \eta_{\zeta_{2}}+12 \eta_{\zeta_{3}}+5 \eta_{\zeta_{5}}\, ,
\end{dmath*}
\begin{dmath*}
 \lambda_{2L}  = 3 \eta_{\zeta_{1}}+3 \eta_{\zeta_{2}}+8 \eta_{\zeta_{3}}+\eta_{\zeta_{4}} \, ,  
 \end{dmath*}
 \begin{dmath*}
  \lambda_{1Y}  = \frac{8}{5} + \frac{49}{5} \eta_{\zeta_{1}}+\frac{49}{5} \eta_{\zeta_{2}}+\frac{24}{5} \eta_{\zeta_{3}}+\frac{3}{5} \eta_{\zeta_{4}}+\frac{64}{5} \eta_{\zeta_{5}}-21 \left(\frac{8}{5} \eta_{PSV}+\frac{6}{5} \eta_{W_R}\right) \, .
\end{dmath*}
\end{dgroup*}

Here again, $PSV$ denotes the Pati-Salam gauge bosons in the $(\overline{3},1,-4)$, $W_R$ the right-handed $W_R^\pm$ in the $(1,1,-6)$.

\begin{table}[ph]
\centering
\begin{center}
\tiny
\begin{tabular}{|c|c|c|c|c|}
\hline
$\bm{SO(10)}$ & $\bm{SU(4)_C \times SU(2)_L \times U(1)_R}$ & $\bm{SU(3)_C \times SU(2)_L \times U(1)_Y}$ & \textbf{Label}  & \textbf{Scale}\\
\hline
\multirow{3}{*}{$10$} & $(6,1,0)$ & & $\zeta_{1}$ & $M_{GUT}$ \\
& $(1,2,1/2)$ & & $\zeta_{2}$ & $M_{GUT}$ \\
\hhline{~----}
& $(1,2,-1/2)$ & $(1,2,-3)$ & H & $M_Z$ \\
\hline
\multirow{7}{*}{$45$} & $(1,1,1)$ & & $\xi_{1}$ & $M_{GUT}$ \\
& $(1,1,0)$ & & $s_{1}$ & $M_{GUT}$ \\
& $(1,1,-1)$ & & $\xi_{2}$ & $M_{GUT}$ \\
& $(1,3,0)$ & & $\zeta_{3}$ & $M_{GUT}$ \\
& $(6,2,1/2)$ & & $\xi_{3}$ & $M_{GUT}$ \\
& $(6,2,-1/2)$ & & $\xi_{4}$ & $M_{GUT}$ \\
& $(15,1,0)$ & & $\zeta_{4}$ & $M_{GUT}$ \\
\hline
\multirow{10}{*}{$120$} & $(1,2,1/2)$ & & $\zeta_{5}$ & $M_{GUT}$ \\
& $(1,2,-1/2)$ & & $\zeta_{6}$ & $M_{GUT}$ \\
& $(10,1,0)$ & & $\zeta_{7}$ & $M_{GUT}$ \\
& $(\overline{10},1,0)$ & & $\zeta_{8}$ & $M_{GUT}$ \\
& $(6,3,1)$ & & $\zeta_{9}$ & $M_{GUT}$ \\
& $(6,1,1)$ & & $\zeta_{10}$ & $M_{GUT}$ \\
& $(6,1,0)$ & & $\zeta_{11}$ & $M_{GUT}$ \\
& $(6,1,-1)$ & & $\zeta_{12}$ & $M_{GUT}$ \\
& $(15,2,1/2)$ & & $\zeta_{13}$ & $M_{GUT}$ \\
& $(15,2,-1/2)$ & & $\zeta_{14}$ & $M_{GUT}$ \\
\hline
\multirow{12}{*}{$\overline{126}$ }& $(\overline{6},1,0)$ & & $\zeta_{15}$ & $M_{GUT}$ \\
& $(\overline{10},3,0)$ & & $\zeta_{16}$ & $M_{GUT}$ \\
\hhline{~----}
& \multirow{3}{*}{$(10,1,1)$}  & $(1,1,0)$ & $s_{2}$ & $M_I$ \\
& & $(3,1,4)$ & $\zeta_{17}$ & $M_I$ \\
& & $(6,1,8)$ & $\xi_{5}$ & $M_I$ \\
\hhline{~----}
& $(10,1,0)$ & & $\zeta_{18}$ & $M_{GUT}$ \\
& $(10,1,-1)$ & & $\zeta_{19}$ & $M_{GUT}$ \\
& $(15,2,1/2)$ & & $\zeta_{20}$ & $M_{GUT}$ \\
\hhline{~----}
& \multirow{4}{*}{$(15,2,-1/2)$}  & $(1,2,-3)$ & $\zeta_{21}$ & $M_I$ \\
& & $(\overline{3},2,-7)$ & $\zeta_{22}$ & $M_I$ \\
& & $(3,2,7)$ & $\zeta_{23}$ & $M_I$ \\
& & $(8,2,-3)$ & $\zeta_{24}$ & $M_I$ \\
\hline
\end{tabular}  
\end{center}
\caption{Decomposition of the scalar representations in an $SO(10)$ model with ${SU(4)_C \times SU(2)_L \times U(1)_R}$ intermediate symmetry. Only relevant decompositions are shown. For further details, see Table~\ref{table:su5decomp}.}
\end{table}

\clearpage

\subsection{$SO(10) \to SU(3)_C \times SU(2)_L \times SU(2)_R \times U(1)_{X}$}

\label{app:so103221thresholdsanddecomposition}

Using Eq.~\eqref{eq:thresholdformula}, we find for the threshold corrections at the $SO(10)$ scale

\begin{dgroup*}
\begin{dmath*}
\lambda_{3C} = 5 -21 \left(4 \eta_{LR}+ \eta_{PSV}\right) + \frac{1}{2} \eta_{\zeta_{1}}+\frac{1}{2} \eta_{\zeta_{2}}+3 \eta_{\zeta_{5}}+\frac{1}{2} \eta_{\zeta_{8}}+\frac{5}{2} \eta_{\zeta_{9}}+\frac{1}{2} \eta_{\zeta_{11}}+\frac{5}{2} \eta_{\zeta_{12}}+\frac{3}{2} \eta_{\zeta_{13}}+\frac{3}{2} \eta_{\zeta_{14}}+\frac{3}{2} \eta_{\zeta_{15}}+\frac{3}{2} \eta_{\zeta_{16}}+2 \eta_{\zeta_{18}}+2 \eta_{\zeta_{19}}+12 \eta_{\zeta_{20}}+\frac{1}{2} \eta_{\zeta_{21}}+\frac{1}{2} \eta_{\zeta_{22}}+\frac{3}{2} \eta_{\zeta_{24}}+\frac{15}{2} \eta_{\zeta_{25}}+\frac{3}{2} \eta_{\zeta_{26}}+\frac{15}{2} \eta_{\zeta_{27}}+2 \eta_{\zeta_{28}}+2 \eta_{\zeta_{29}}+12 \eta_{\zeta_{30}} \, , 
\end{dmath*}
\begin{dmath*}
 \lambda_{2L}  = 6 -21 (3 \eta_{V_{1}}+3 \eta_{V_{2}}) + 2 \eta_{\zeta_{4}}+\eta_{\zeta_{6}}+6 \eta_{\zeta_{13}}+6 \eta_{\zeta_{14}}+\eta_{\zeta_{17}}+3 \eta_{\zeta_{18}}+3 \eta_{\zeta_{19}}+8 \eta_{\zeta_{20}}+2 \eta_{\zeta_{23}}+6 \eta_{\zeta_{24}}+12 \eta_{\zeta_{25}}+3 \eta_{\zeta_{28}}+3 \eta_{\zeta_{29}}+8 \eta_{\zeta_{30}} \, ,  
 \end{dmath*}
 \begin{dmath*}
 \lambda_{2R}  = 6-21 (3 \eta_{V_{1}}+3 \eta_{V_{2}}) + 2 \eta_{\zeta_{3}}+\eta_{\zeta_{6}}+6 \eta_{\zeta_{15}}+6 \eta_{\zeta_{16}}+\eta_{\zeta_{17}}+3 \eta_{\zeta_{18}}+3 \eta_{\zeta_{19}}+8 \eta_{\zeta_{20}}+6 \eta_{\zeta_{26}}+12 \eta_{\zeta_{27}}+3 \eta_{\zeta_{28}}+3 \eta_{\zeta_{29}}+8 \eta_{\zeta_{30}} \, ,  
 \end{dmath*}
 \begin{dmath*}
  \lambda_{1X}  = 8 -21 (4 \eta_{LR}+4 \eta_{PSV}) + \frac{1}{2} \eta_{\zeta_{1}}+\frac{1}{2} \eta_{\zeta_{2}}+\frac{3}{2} \eta_{\zeta_{7}}+\frac{1}{2} \eta_{\zeta_{8}}+\eta_{\zeta_{9}}+\frac{3}{2} \eta_{\zeta_{10}}+\frac{1}{2} \eta_{\zeta_{11}}+\eta_{\zeta_{12}}+\frac{3}{2} \eta_{\zeta_{13}}+\frac{3}{2} \eta_{\zeta_{14}}+\frac{3}{2} \eta_{\zeta_{15}}+\frac{3}{2} \eta_{\zeta_{16}}+8 \eta_{\zeta_{18}}+8 \eta_{\zeta_{19}}+\frac{1}{2} \eta_{\zeta_{21}}+\frac{1}{2} \eta_{\zeta_{22}}+\frac{9}{2} \eta_{\zeta_{23}}+\frac{3}{2} \eta_{\zeta_{24}}+3 \eta_{\zeta_{25}}+\frac{3}{2} \eta_{\zeta_{26}}+3 \eta_{\zeta_{27}}+8 \eta_{\zeta_{28}}+8 \eta_{\zeta_{29}} \, 
\end{dmath*}
\end{dgroup*}

and for the corrections at the $SU(3)_C \times SU(2)_L \times SU(2)_R \times U(1)_X$ scale

\begin{dgroup*}
\begin{dmath*}
\lambda_{3C} = 5 \, ,
\end{dmath*}
\begin{dmath*}
 \lambda_{2L}  = 6+ \eta_{\zeta_{1}}+\eta_{\zeta_{2}}+\eta_{\zeta_{3}}\, ,  
 \end{dmath*}
 \begin{dmath*}
  \lambda_{1Y}  = 8+ \frac{3}{5} \eta_{\zeta_{1}}+\frac{3}{5} \eta_{\zeta_{2}}+\frac{3}{5} \eta_{\zeta_{3}} \, .
\end{dmath*}
\end{dgroup*}

Here, $PSV$ denotes the Pati-Salam gauge bosons in the $(\overline{3},1,1,-4/3)$ representation and $LR$ the additional bosons in the $(3,2,2,-2/3)$.

\renewcommand\arraystretch{1.5}
\begin{table}[ph]
\centering
\begin{center}
\tiny
\begin{tabular}{|c|c|c|c|c|}
\hline
$\bm{SO(10)}$ & $\bm{SU(3)_C \times SU(2)_L \times SU(2)_R \times U(1)_{X}}$ & $\bm{SU(3)_C \times SU(2)_L \times U(1)_Y}$ & \textbf{Label}  & \textbf{Scale}\\
\hline
\multirow{3}{*}{$10$} & $(3,1,1,-2/3)$ & & $\Omega_{1}$ & $M_U$ \\
& $(\overline{3},1,1,2/3)$ & & $\Omega_{2}$ & $M_U$ \\
\hhline{~----}
& \multirow{2}{*}{$(1,2,2,0)$}  & $(1,2,3)$ & $\Omega_{3}$ & $M_I$ \\
& & $(1,2,-3)$ & $H$ & $M_Z$ \\
\hline
\multirow{8}{*}{$45$} & $(1,1,3,0)$ & & $\Omega_{4}$ & $M_U$ \\
& $(1,3,1,0)$ & & $\Omega_{5}$ & $M_U$ \\
& $(3,2,2,-2/3)$ & & $\xi_{1}$ & $M_U$ \\
& $(\overline{3},2,2,2/3)$ & & $\xi_{2}$ & $M_U$ \\
& $(1,1,1,0)$ & & $s_{1}$ & $M_U$ \\
& $(3,1,1,4/3)$ & & $\xi_{3}$ & $M_U$ \\
& $(\overline{3},1,1,-4/3)$ & & $\xi_{4}$ & $M_U$ \\
& $(8,1,1,0)$ & & $\Omega_{6}$ & $M_U$ \\
\hline
\multirow{14}{*}{$120$} & $(1,2,2,0)$ & & $\Omega_{7}$ & $M_U$ \\
& $(1,1,1,2)$ & & $\Omega_{8}$ & $M_U$ \\
& $(3,1,1,2/3)$ & & $\Omega_{9}$ & $M_U$ \\
& $(6,1,1,-2/3)$ & & $\Omega_{10}$ & $M_U$ \\
& $(1,1,1,-2)$ & & $\Omega_{11}$ & $M_U$ \\
& $(\overline{3},1,1,-2/3)$ & & $\Omega_{12}$ & $M_U$ \\
& $(\overline{6},1,1,2/3)$ & & $\Omega_{13}$ & $M_U$ \\
& $(3,3,1,2/3)$ & & $\Omega_{14}$ & $M_U$ \\
& $(\overline{3},3,1,-2/3)$ & & $\Omega_{15}$ & $M_U$ \\
& $(3,1,3,2/3)$ & & $\Omega_{16}$ & $M_U$ \\
& $(\overline{3},1,3,-2/3)$ & & $\Omega_{17}$ & $M_U$ \\
& $(1,2,2,0)$ & & $\Omega_{18}$ & $M_U$ \\
& $(\overline{3},2,2,-4/3)$ & & $\Omega_{19}$ & $M_U$ \\
& $(3,2,2,4/3)$ & & $\Omega_{20}$ & $M_U$ \\
& $(8,2,2,0)$ & & $\Omega_{21}$ & $M_U$ \\
\hline
\multirow{14}{*}{$\overline{126}$} & $(3,1,1,-2/3)$ & & $\Omega_{22}$ & $M_U$ \\
& $(\overline{3},1,1,2/3)$ & & $\Omega_{23}$ & $M_U$ \\
& $(1,3,1,2)$ & & $\Omega_{24}$ & $M_U$ \\
& $(\overline{3},3,1,2/3)$ & & $\Omega_{25}$ & $M_U$ \\
& $(\overline{6},3,1,-2/3)$ & & $\Omega_{26}$ & $M_U$ \\
& $(1,1,3,-2)$ & & $s_{2}$ & $M_I$ \\
& $(3,1,3,-2/3)$ & & $\Omega_{27}$ & $M_U$ \\
& $(6,1,3,2/3)$ & & $\Omega_{28}$ & $M_U$ \\
\hhline{~----}
& \multirow{2}{*}{$(1,2,2,0)$}  & $(1,2,3)$ & $\Omega_{29}$ & $M_I$ \\
& & $(1,2,-3)$ & $\Omega_{30}$ & $M_I$ \\
\hhline{~----}
& $(3,2,2,4/3)$ & & $\Omega_{31}$ & $M_U$ \\
& $(\overline{3},2,2,-4/3)$ & & $\Omega_{32}$ & $M_U$ \\
& $(8,2,2,0)$ & & $\Omega_{33}$ & $M_U$ \\
\hline
\end{tabular}  
\end{center}
\caption{Decomposition of the scalar representations in an $SO(10)$ model with ${SU(3)_C \times SU(2)_L \times SU(2)_R \times U(1)_{X}}$ intermediate symmetry. Only relevant decompositions are shown. For further details, see Table~\ref{table:su5decomp}.}
\end{table}

\clearpage

\bibliographystyle{h-physrev}
\bibliography{bib}

\end{document}